%% file: LH_StandardModel_Report.tex
\pdfoutput=1

\documentclass[12pt,letterpaper,hyperref,dvipsnames]{cernyrep}
\usepackage[utf8]{inputenc}

\hypersetup{
  colorlinks=true,
  citecolor=blue,
  linkcolor=red,
  urlcolor=blue,
  pdfauthor={LH23 participants},
  pdftitle={Les Houches 2023 Proceedings}
}

\usepackage{cite}
\usepackage{amsmath,amssymb,graphicx,xspace,subfigure}
\usepackage{mathtools}
\usepackage{wasysym}
\usepackage{todonotes}
\usepackage{bigints}
\usepackage{wrapfig}

\usepackage[T1]{fontenc}
\usepackage{lmodern}
\usepackage{rotating}
\usepackage{xcolor}
\usepackage{booktabs}
\usepackage{arydshln}
\usepackage{lscape}
\usepackage{floatpag}
\floatpagestyle{plain}
\usepackage{axodraw}
\usepackage{lineno}
\usepackage{graphpap}

\usepackage{cernunits}

\usepackage{import}
\usepackage{enumitem}
\usepackage{caption}
\usepackage{xspace}
\usepackage{adjustbox}
\usepackage{tikz}
\usetikzlibrary{positioning,arrows}
\usetikzlibrary{decorations.pathmorphing}
\usetikzlibrary{decorations.markings}

\input{main_macros}

\begin{document}



\begin{center}
{\large {\bf LES HOUCHES 2023: PHYSICS AT TEV COLLIDERS \\[4mm]}}
{\Large {\bf STANDARD MODEL WORKING GROUP REPORT}}
\end{center}

\pagenumbering{roman}

~\\~
\input{conveners}

\noindent{{\bf Abstract}}\\[.2cm]
This report presents a short summary of the activities of the ``Standard Model'' working group for the ``Physics at TeV Colliders'' workshop (Les Houches, France, 12--30 June, 2023). 
~\\~\\~

\noindent{\bf Acknowledgements}\\[.2cm]
We would like to thank
the Les Houches staff for the stimulating environment always present at Les Houches. We thank the Formation Permanente of CNRS, the IDEX Université Grenoble Alpes, the Labex Enigmass, the Université Savoie Mont Blanc, LAPP and LAPTh for support.

\newpage
\begingroup
\let\clearpage\relax
\include{all_authors}

~\\~
\include{all_affiliations}

\endgroup




\newpage
{\parskip=1.5ex \tableofcontents}

\newpage
\pagenumbering{arabic}
\setcounter{footnote}{0}

\chapter*{\Large Introduction}
\addcontentsline{toc}{chapter}{\Large Introduction}
\begingroup
\let\clearpage\relax
\subimport{introduction}{intro}
\endgroup

\newpage
\chapter{\Large Higgs Physics and SM phenomenology}
\label{cha:sm}
\input{sm_pheno/sm_pheno.main}

\newpage
\chapter{\Large Monte Carlo generators, Tools, Machine-Learning}
\label{cha:mc}
\input{tools/tools.main}


\clearpage

\bibliography{LH23_SM_Biblio}

\end{document}

%% file: main_macros.tex
\makeatletter
\renewcommand{\thechapter}{\@Roman\c@chapter}
\makeatother

\newcommand{\pt}{$p_T$}
\newcommand{\ee}{$ee$}
\newcommand{\pp}{$ep$}
\newcommand{\herwig}{\textsc{Herwig}\xspace}
\newcommand{\herwigpp}{\textsc{Herwig++}\xspace}
\newcommand{\pythia}{\textsc{Pythia}\xspace}
\newcommand{\sherpa}{\textsc{Sherpa}\xspace}
\newcommand{\powhegbox}{\textsc{Powheg Box}\xspace}
\newcommand{\powheg}{\textsc{Powheg}\xspace}
\newcommand{\mgamcnlo}{\textsc{MadGraph5\_aMC@NLO}\xspace}
\newcommand{\mgamc}{\textsc{MG5aMC}\xspace}
\newcommand{\chilipepper}{\textsc{Chili/Pepper}\xspace}
\newcommand{\madgraph}{\textsc{MadGraph}\xspace}

\newcommand{\ttbb}{$t\bar{t}b\bar{b}$~}
\newcommand{\GeV}{GeV}


%% file: conveners.tex
\noindent \textbf{Conveners}


\noindent \textit{Higgs Physics:}\\
\noindent M.~Donegà (CMS), 
S.~Jones (Theory), 
K.~Köneke (ATLAS), 
R.~Röntsch (Theory)
~\\~\\
\noindent \textit{Standard Model Phenomenology:} \\
\noindent P.~Azzurri (CMS),
A.~Hinzmann (CMS, Jet-Substructure subgroup),
A.~Huss (Theory),
J.~Huston (ATLAS),
S.~Marzani (Theory, Jet-Substructure subgroup),	
M.~Pellen (Theory)
~\\~\\
\noindent \textit{Monte Carlo generators, Tools, Machine-Learning:}\\ 
\noindent S.~Höche (Theory),
J.~McFayden (ATLAS),
V.~Mikuni (ML contact),
S.~Plätzer (Theory)
~\\~\\

%% file: all_authors.tex
\noindent \textbf{Authors and Participants}

\noindent
J.~Andersen$^{1}$,              
B.~Assi$^{2}$,                  
K.~Asteriadis$^{3}$,            
P.~Azzurri$^{4a}$,              
G.~Barone$^{5}$,                
A.~Behring$^{6}$,               
A.~Benecke$^{7}$,               
S.~Bhattacharya$^{8}$,          
E.~Bothmann$^{9}$,             
S.~Caletti$^{10}$,              
X.~Chen$^{11}$,                 
M.~Chiesa$^{12}$,               
A.~Cooper-Sarkar$^{13}$,        
T.~Cridge$^{14}$,               
A.~Cueto Gomez$^{15}$,          
S.~Datta$^{16}$,                
P.~K.~Dhani$^{17}$,             
M.~Donegà$^{18}$,               
T.~Engel$^{19}$,                
S.~Ferrario Ravasio$^{6}$,      
S.~Forte$^{20a,20b}$,            
P.~Francavilla$^{4a,4b}$,       
M.V.~Garzelli$^{21}$,          
A.~Ghira$^{22a,22b}$,           
A.~Ghosh$^{23,24}$,             
F.~Giuli$^{25a,25b}$,           
L.~Gouskos$^{5}$,              
P.~Gras$^{26}$,                
C.~G{\"u}tschow$^{27}$,        
Y.~Haddad$^{28}$,              
L.~Harland-Lang$^{27}$,        
F. Hekhorn$^{29,20a,20b}$,       
I.~Helenius$^{29}$,            
A.~Hinzmann$^{14}$,            
S.~H{\"o}che$^{2}$,            
J.~Holguin$^{30,31}$,          
A.~Huss$^{6}$,                 
J.~Huston$^{32}$,              
T.~Je{\v z}o$^{33}$,           
S.~Jones$^{1}$,                
S.~Kiebacher$^{34}$,           
M.~Knobbe$^{9}$,              
R.~Kogler$^{14}$,              
K.~Köneke$^{19}$,              
L.~Kunz$^{34}$,                
M.~LeBlanc$^{5,30}$,           
P.~Loch$^{35}$,                
G.~Loeschcke Centeno$^{36}$,   
M.~Löschner$^{14}$,            
A.~Maas$^{37}$,                
G.~Magni$^{38}$,               
A.~Maier$^{14}$,               
M.~Marcoli$^{1}$,              
S.~Marzani$^{22a,22b}$,         
J.~McFayden$^{36}$,            
P.~Meinzinger$^{1}$,           
V.~Mikuni$^{39}$,              
S.~Moch$^{21}$,                
P.~Nadolsky$^{40}$,            
D.~Napoletano$^{41a,41b}$,      
M.~Pellen$^{19}$,              
S.~Plätzer$^{37,42}$,          
R.~Poncelet$^{43}$,            
C.~Preuss$^{44}$,              
H.~Qu$^{45}$,                  
K.~Rabbertz$^{46}$,            
D.~Reichelt$^{1}$,             
A.~Rescia$^{14,22a}$,          
J.~Roloff$^{5}$,               
R.~Röntsch$^{20a,20b}$,         
S.~Sanchez Cruz$^{45}$,        
T.~Sarkar$^{4a}$,              
L.~Scyboz$^{47,48}$,           
F.~Sforza$^{22a,22b}$,          
A.~Siódmok$^{49}$,             
G.~Stagnitto$^{41a,41b}$,       
A.~Tarek$^{32}$,               
R.S.~Thorne$^{27}$,            
A.~Valassi$^{50}$,             
J.~Whitehead$^{43}$,           
J.~Winter$^{51}$               
~\\~\\~
\noindent \textbf{Organizers}

\noindent
{C.~Delaunay}$^{52}$, 
{B.~Herrmann}$^{52}$,
{E.~Re}$^{41a,41b,52}$

%% file: all_affiliations.tex
\begin{itemize}
\vspace*{-.3cm}\item[$^{1}$] Institute for Particle Physics Phenomenology, Durham University, Durham, UK
\vspace*{-.3cm}\item[$^{2}$] Fermi National Accelerator Laboratory, Batavia, IL, USA
\vspace*{-.3cm}\item[$^{3}$] Universität Regensburg, Regensburg, Germany
\vspace*{-.3cm}\item[$^{4a}$] INFN, Sezione di Pisa, Pisa, Italy
\vspace*{-.3cm}\item[$^{4b}$] Dipartimento di Fisica E. Fermi, Università di Pisa, Pisa, Italy
\vspace*{-.3cm}\item[$^{5}$] Brown University, Providence, RI, USA
\vspace*{-.3cm}\item[$^{6}$] Theoretical Physics Department, CERN, Geneva, Switzerland
\vspace*{-.3cm}\item[$^{7}$] Centre for Cosmology, Particle Physics and Phenomenology, Université catholique de Louvain, Louvain-la-Neuve, Belgium
\vspace*{-.3cm}\item[$^{8}$] Wayne State University, Detroit, MI, USA
\vspace*{-.3cm}\item[$^{9}$] Institut für Theoretische Physik, Georg-August-Universität Göttingen, Göttingen, Germany
\vspace*{-.3cm}\item[$^{10}$] Institute for Theoretical Physics, ETH, Zürich, Switzerland
\vspace*{-.3cm}\item[$^{11}$] School of Physics, Shandong University, Jinan, Shandong, China 
\vspace*{-.3cm}\item[$^{12}$] INFN, Sezione di Pavia, Pavia, Italy
\vspace*{-.3cm}\item[$^{13}$] Department of Physics, University of Oxford, Oxford, UK
\vspace*{-.3cm}\item[$^{14}$] Deutsches Elektronen-Synchrotron DESY, Germany
\vspace*{-.3cm}\item[$^{15}$] Departamento de F\'isica Teorica C-15 and CIAFF, Universidad Aut\'onoma de Madrid, Madrid, Spain
\vspace*{-.3cm}\item[$^{16}$] Centre for High Energy Physics, Indian Institute of Science, Bangalore, India
\vspace*{-.3cm}\item[$^{17}$] IFIC, University of Valencia - CSIC, Valencia, Spain
\vspace*{-.3cm}\item[$^{18}$] Institute for Particle Physics and Astrophysics, ETH, Zürich, Switzerland
\vspace*{-.3cm}\item[$^{19}$] Physikalisches Institut, Albert-Ludwigs-Universität Freiburg, Freiburg, Germany
\vspace*{-.3cm}\item[$^{20a}$] Tif Lab, Dipartimento di Fisica, Università di Milano, Milano, Italy
\vspace*{-.3cm}\item[$^{20b}$] INFN, Sezione di Milano, Milano, Italy
\vspace*{-.3cm}\item[$^{21}$] II. Institute for Theoretical Physics, Hamburg University, Hamburg, Germany
\vspace*{-.3cm}\item[$^{22a}$] Dipartimento di Fisica, Università di Genova, Genova, Italy
\vspace*{-.3cm}\item[$^{22b}$] INFN, Sezione di Genova, Genova, Italy
\vspace*{-.3cm}\item[$^{23}$] Department of Physics and Astronomy, University of California, Irvine, CA, USA
\vspace*{-.3cm}\item[$^{24}$] Physics Division, Lawrence Berkeley National Laboratory, CA, USA
\vspace*{-.3cm}\item[$^{25a}$] Dipartimento di Fisica, Università di Roma Tor Vergata, Roma, Italy
\vspace*{-.3cm}\item[$^{25b}$] INFN, Sezione di Roma2, Roma, Italy
\vspace*{-.3cm}\item[$^{26}$] IRFU, CEA, Université Paris-Saclay, Gif-sur-Yvette, France
\vspace*{-.3cm}\item[$^{27}$] Department of Physics and Astronomy, University College London, London, UK
\vspace*{-.3cm}\item[$^{28}$] Northeastern University, Boston, MA, USA
\vspace*{-.3cm}\item[$^{29}$] Department of Physics, University of Jyvaskyla, Finland, and Helsinki Institute of Physics, University of Helsinki, Finland
\vspace*{-.3cm}\item[$^{30}$] School of Physics and Astronomy, University of Manchester, Manchester, UK
\vspace*{-.3cm}\item[$^{31}$] CPHT, CNRS, École polytechnique, Institut Polytechnique de Paris, Palaiseau, France
\vspace*{-.3cm}\item[$^{32}$] Department of Physics and Astronomy, Michigan State University, East Lansing, MI, USA
\vspace*{-.3cm}\item[$^{33}$] Institute of Theoretical Physics, Universität Münster, Münster, Germany
\vspace*{-.3cm}\item[$^{34}$] Institute for Theoretical Physics, KIT, Karlsruhe, Germany
\vspace*{-.3cm}\item[$^{35}$] Department of Physics, University of Arizona, Tucson, AZ, USA
\vspace*{-.3cm}\item[$^{36}$] Department of Physics and Astronomy, University of Sussex, Brighton, UK
\vspace*{-.3cm}\item[$^{37}$] Institute of Physics, NAWI Graz, University of Graz, Graz, Austria
\vspace*{-.3cm}\item[$^{38}$] Department of Physics and Astronomy, Vrije Universiteit and Nikhef Theory Group, Amsterdam, The Netherlands
\vspace*{-.3cm}\item[$^{39}$] National Energy Research Scientific Computing Center (NERSC), Berkeley, CA, USA
\vspace*{-.3cm}\item[$^{40}$] Department of Physics, Southern Methodist University, Dallas, TX, USA
\vspace*{-.3cm}\item[$^{41a}$] Dipartimento di Fisica, Università di Milano-Bicocca, Milano, Italy
\vspace*{-.3cm}\item[$^{41b}$] INFN, Sezione di Milano-Bicocca, Milano, Italy
\vspace*{-.3cm}\item[$^{42}$] Particle Physics, Faculty of Physics, University of Vienna, Wien, Austria
\vspace*{-.3cm}\item[$^{43}$] The Henryk Niewodniczański Institute of Nuclear Physics, Polish Academy of Sciences, Krakow, Poland
\vspace*{-.3cm}\item[$^{44}$] Department of Physics, University of Wuppertal, Wuppertal, Germany
\vspace*{-.3cm}\item[$^{45}$] Experimental Physics Department, CERN, Geneva, Switzerland
\vspace*{-.3cm}\item[$^{46}$] Institute for Experimental Physics, KIT, Karlsruhe, Germany
\vspace*{-.3cm}\item[$^{47}$] School of Physics and Astronomy, Monash University, Clayton, Australia
\vspace*{-.3cm}\item[$^{48}$] Rudolf Peierls Centre for Theoretical Physics, Department of Physics, University of Oxford, Oxford, UK
\vspace*{-.3cm}\item[$^{49}$] Jagiellonian University, Krakow, Poland
\vspace*{-.3cm}\item[$^{50}$] IT Department, CERN, Geneva, Switzerland
\vspace*{-.3cm}\item[$^{51}$] University at Buffalo, The State University of New York, Buffalo, NY, USA
\vspace*{-.3cm}\item[$^{52}$] Laboratoire d’Annecy-le-Vieux de Physique Théorique, CNRS and Université Savoie Mont Blanc, Annecy, France
\end{itemize}

This research was supported by the Fermi National Accelerator
Laboratory (Fermilab), a U.S.\ Department of Energy, Office of
Science, HEP User Facility.  Fermilab is managed by Fermi Research
Alliance, LLC (FRA), acting under Contract No. DE--AC02--07CH11359.
This research used resources of the National Energy Research
Scientific Computing Center (NERSC), a U.S. Department of Energy
Office of Science User Facility. E.B. and M.K. acknowledge support
from BMBF (contract 05H21MGCAB).  Their research is funded by the
Deutsche Forschungsgemeinschaft (DFG, German Research Foundation) -
456104544; 510810461.  This work used computing resources of the Emmy
HPC system provided by The North-German Supercomputing Alliance
(HLRN).  T.C acknowledges funding from the European Research Council
(ERC) under the European Union’s Horizon 2020 research and innovation
programme (Grant agreement No. 101002090 COLORFREE). This work was
supported by grant 2019/34/E/ST2/00457 of the National Science
Centre,724 Poland.  A.S. is also supported by the Priority Research
Area Digiworld under the program ``Excellence Initiative - Research
University'' at the Jagiellonian University in Krakow. The work of
J.M. was supported by the Royal Society fellowship
grant URF\textbackslash R1\textbackslash 201519.

%% file: introduction/intro.tex
This document is the report of the Standard Model Physics session of
the 2023 Les Houches Workshop ``Physics at TeV Colliders'' (PhysTeV). The workshop
brought together theorists and experimenters who discussed a
significant number of novel ideas related to physics at colliders, the
main focus being precision phenomenology.

New computational methods and techniques, including Machine Learning
(ML), were considered, with the aim of improving the technology
available for theoretical, phenomenological and experimental physics
studies at colliders, particularly in the context of Standard Model
predictions. For the 2023 edition, the topics to be discussed during
the workshop were subdivided in 3 main working groups (Higgs Physics
-- Standard Model Phenomenology -- Monte Carlo generators, Tools, and
Machine Learning), with a sub-group dedicated to theoretical and
phenomenological aspects of jets and Jet-Substructure (JSS) Techniques
at colliders.

A compact summary of the topics discussed during the workshop and of
the ongoing studies is reported in the rest of this introductory
section. In chapter
\ref{cha:sm} and \ref{cha:mc} more details are discussed, without a strict
distinction between working groups. Several discussions and ideas for
future studies involved a large number of participants. In some cases,
where appropriate, a detailed list of contributors is instead
included.

Further developments, where necessary, will be included in future
revisions of this document, as well as explicit references to
published papers and completed studies that originated from the 2023
PhysTeV workshop.
~\\~\\~

A traditional output of the ``SM session'' of the PhysTeV workshop has
been the compilation of a ``wishlist'' of precision computations that
are particularly important to complete in order to optimize the
physics information that can be extracted from experimental
measurements at the LHC. Discussions took place during the workshop,
and an update of such list, with the currently more pressing
experimental needs, is planned. A few highlights are presented in this document. 

In the past few years, it has been possible to extract important
information about N3LO splitting functions, thereby allowing the production of
PDFs with approximate N3LO evolution. Although only an exact N3LO
computation of the splitting functions will allow fully consistent N3LO
computations for cross sections at colliders, the aforementioned
information 
already allows for a number of useful 
studies, such as benchmarking N3LO evolution among different PDF
groups, or assessing the impact of approximate N3LO PDFs on
uncertainties in the computation of gluon fusion Higgs production.

Vector boson fusion (VBF) and vector boson scattering (VBS) processes
have been discussed in different aspects, with both theoretical and
experimental inputs; on the one hand, starting from discussions that
took place previously, for example within the LHC EW working group, a
possible set of recommendations for future VBF and VBF measurements
was proposed, with the aim of moving towards common definitions of
cross sections (and fiducial cuts) among different experimental
collaborations, that allows for clean theoretical
predictions. Workshop participants have also made progress on an
ongoing effort of the LHCHXSWG whose aim is to outline the definitions
of differential predictions to be obtained at fixed order for VBF, and
subsequently to check the impact of uncertainties due to parton
showers. One of the largest uncertainties in the study of VBF Higgs
boson production is the background from gluon-gluon fusion Higgs + 2
jet production. Following up on a previous Les Houches, a detailed
study of this source of background is being carried out, and is
briefly summarized in a one-pager in these proceedings.  There were
also preliminary discussions about modifying the Simplified Template
Cross Sections (STXS) current binnings (possibly including also other
variables) in order to improve the sensitivity to the Higgs
self-coupling in $t\bar{t}H$ production.

In view of the ever-increasing precision of experimental measurements,
not only is the inclusion of (at least) NLO electroweak corrections
necessary from a phenomenological point of view, but also a
quantification of the associated theoretical uncertainties becomes
potentially relevant. Existing prescriptions were discussed in Les
Houches, and have been summarized in these proceedings.

In the year preceding the workshop, several theoretical ideas to
properly define the flavour of jets at colliders have been put
forward. This edition of the workshop in Les Houches was suitable to
allow for a timely detailed comparison among such methods, in
different processes, with different theoretical predictions
(fixed-order vs. complete Monte Carlo simulations), and with a view on
experimental feasibility. Such comprehensive study is currently underway, and a short summary is included in these proceedings.

Besides the impact expected from the major progress towards
NLL-accurate (and beyond) parton showers, ideas have been put forward
to better understand differences among event generators, with the
ultimate goal of improving the way uncertainties due to Monte Carlo are
typically quantified: examples range from a more accurate study of the
interplay between parton showering stage and hadronization, to the use
of differential measurements of jet substructure observables and of LEP data. On
the former aspect, for example, work is in progress to identify hadron
collider observables particularly sensitive to such stages of the
simulations, both by means of a survey of existing measurements, as
well as by the development of new observables.
There is also an ongoing effort to perform a survey on existing jet-
and JSS-measurements in order to reveal the built-up of the jet
structure by means of correlations among jet constituents.

In Les Houches an effort has been initiated in order to check to which
extent tools are ready for describing photoproduction at an
electron-ion collider.

There were also discussions on the status and future prospects for the
transition of numerical codes from CPU to GPU. The possibility to
perform a benchmarking among the two tools that can already compute
amplitudes on GPU has been raised. The workshop was also useful to
discuss plans for future developments in this area, such as going
beyond LO, or porting to GPUs tools where this transition has not
started yet.

The role of Machine-Learning techniques in modern phenomenology is now
well established. In Les Houches there were several discussions on
this topic. As far as unbinned unfolding is concerned, experts have
discussed both recommendations for its practical use, as well as
future prospects. In a different context, regarding analysis
techniques that aim to measure physical parameters, the use of the
so-called \emph{Weight Derivative Regression} was discussed, which
could also be used efficiently to optimize analysis methods using ML.
ML is also being actively studied to tag longitudinally-polarised
events on an event-by-event basis, and in Les Houches other processes
where similar techniques can be applied were discussed. More details
can be found in the dedicated sections.

As per tradition, the PhysTeV workshop often has proven to be an
useful place to discuss technical and computational improvements in
state-of-the-art computations and event simulations. In the 2023
edition, novel strategies to improve the simulations plagued by a
large fraction of negative-weighted events were tested and compared:
a) the \emph{cell resampler} technique was applied to reduce the
negative-weighted events within an event sample similar to the one
used by experimental collaborations to estimate backrounds to $pp\to\
H\to \gamma\gamma$; b) a comparison of the efficiency
between \emph{cell resampling} and a ML-based approach (\emph{neural
positive reweighting}) was performed on a \ttbb sample.
There were also discussions on the need to update accords and
interfaces between different tools, especially in light of recent
developments on parton showers algorithms.



%% file: sm_pheno/sm_pheno.main.tex

\begingroup
\let\clearpage\relax
\subimport{sm_pheno/wishlist/}{wishist_main}
\subimport{sm_pheno/N3LO_benchmark/}{N3LO_benchmark_main}
\subimport{sm_pheno/N3LOggH/}{N3LOggH_main}
\subimport{sm_pheno/H2J/}{H2J_main}
\subimport{sm_pheno/VBFVBS/}{vbfvbs_main}
\subimport{sm_pheno/VBF_hxswg/}{VBF_hxswg_main}
\subimport{sm_pheno/EW/}{EW_main}
\subimport{sm_pheno/flavouredjets/}{flavouredjets_main}
\subimport{sm_pheno/ttH_STXS/}{ttH_STXS}
\nopagebreak
\endgroup

%% file: sm_pheno/wishlist/wishist_main.tex
\section{The 2023 Les Houches Wishlist}
\emph{\noindent A.~Huss, J.~Huston, S.~Jones, M.~Pellen,  R. R{\"o}ntsch}
\vspace*{0.5em}                                 

\noindent It is clear that the presence of new physics at the LHC is doing a good job of hiding. Barring any unexpected thresholds crossed with the amount of data to be taken in the HL-LHC, the key to both understanding the SM physics we are measuring at the LHC, and the use of that physics to put constraints on/discover BSM physics requires improvements in theoretical calculations and in experimental analysis techniques.  The rate of progress in higher order QCD and EW calculations is such that an update is necessary to keep track of the ongoing efforts. The Les Houches wishlist for precision calculations was started for just such a purpose and has been going on for over a decade, serving as a successor to the Les Houches NLO wishlist~\cite{Buttar:2006zd}. 
Indeed, although Les Houches was not physically held in 2021, it was felt necessary to provide an update to the wishlist~\cite{Huss:2022ful}. A complete update of the wishlist is in progress, and will be posted to the archive when ready. This report is a one-pager serving  as a brief summary/preview of that effort. 

In terms of multiloop amplitude calculations, the frontier has now moved on from $2\rightarrow2$ processes at NNLO to $2\rightarrow3$, and to $2\rightarrow2$ at N3LO.  Recent calculations for the former include 3 photon production~\cite{Chawdhry:2019bji,Abreu:2020cwb,Chawdhry:2020for,Kallweit:2020gcp,Abreu:2023bdp}, diphoton-plus-jet production~\cite{Chawdhry:2021hkp,Badger:2021ohm,Badger:2021imn}, photon + 2 jet production~\cite{Badger:2023mgf} and 3 jet production~\cite{Abreu:2018jgq,Abreu:2021oya,Czakon:2021mjy,DeLaurentis:2023nss,Agarwal:2023suw,DeLaurentis:2023izi} at NNLO, which very recently became available also at full color. Amplitude calculations for $2\rightarrow2$ at 3-loops include four-parton scattering~\cite{Caola:2021rqz,Caola:2021izf,Caola:2022dfa}, diphoton production~\cite{Bargiela:2021wuy}, and $V+j$ production~\cite{Gehrmann:2023jyv}. The approximate $t\bar{t}H$ cross section has been calculated to NNLO, with the two-loop virtual amplitudes computed using the approximation that the Higgs $p_T$~\cite{Catani:2022mfv} is small. Although this approximation should be adequate numerically, it is nevertheless still desirable for these two-loop amplitudes to be calculated exactly. There has been recent progress in this direction~\cite{Agarwal:2024jyq,Wang:2024pmv,Buccioni:2023okz,FebresCordero:2023gjh}. The frontier for $2\rightarrow3$ processes includes critical processes such as Hjj, to serve as a better calculation of the ggF backgrounds to VBF Higgs production. The 5-point, one mass, master integrals relevant for V+2j and H+2j have now been calculated~\cite{Chicherin:2021dyp,Abreu:2023rco}. The computation of 5-point, two mass, master integrals is also advancing rapidly, with recent results for the leading colour of $t \overline{t}+j$~\cite{Badger:2022hno,Badger:2024fgb}.

Looking forward in the Higgs sector, there has been a great deal of progress in computing the three-loop amplitudes required for the prediction of  $H+j$ production at N3LO~\cite{Henn:2023vbd} in the heavy-top limit (HTL), and even inclusive Higgs boson production at N4LO in the HTL might be possible in the next few years~\cite{Lee:2022nhh,Lee:2023dtc}. 
However, one should bear in mind that the treatment of IR singularities at N3LO for processes involving a jet has not yet been developed, and is expected to provide a significant obstacle to the computation of $H+j$ production at this order. In this context, there might also be issues with numerical stability arising from e.g.\ the need to compute two-loop amplitudes involving a Higgs and four partons, in the limit where one of the final-state partons becomes unresolved.  

Progress in the development of methods to treat infrared singularities have mostly focused on refining and generalizing methods for NNLO computations, with a particular focus on processes with high partonic multiplicity. 
In the context of the antenna subtractions, there have been efforts to streamline the construction of antenna functions by reducing the number of spurious sublimits~\cite{Braun-White:2023sgd,Braun-White:2023zwd,Fox:2023bma} as well as the formulation of the method in color space~\cite{Gehrmann:2023dxm,Chen:2022ktf} allowing high-multiplicity processes to be computed beyond the leading-color approximation. Parallel efforts in the local analytic subtraction methods have succeeded in demonstrating the pole cancellation in fully differential observables in the production of arbitrarily many massless partons in $e^-e^+$ collisions~\cite{Bertolotti:2022aih}, with an extension to hadroproduction processes underway (see Ref.~\cite{Bertolotti:2022ohq}). Similarly, pole cancellation has been demonstrated for the production of arbitrarily many gluons in $q\bar{q}$ annihilation in the nested soft-collinear subtraction scheme~\cite{Devoto:2023rpv}, anticipating an extension to arbitrary hadronic partonic processes quite soon. 
Apart from these recent developments, we also point out that the STRIPPER method has proven to be sufficiently flexible to handle the subtractions required for trijet production at NNLO. The outlook for NNLO subtraction methods is therefore very positive: it is not unreasonable to imagine that, in the next few years, several different subtraction methods will be available to treat high-multiplicity production processes at NNLO. Looking to the next order, results for $e^+e^- \to jj$ have been presented using antenna subtraction in Ref.~\cite{Jakubcik:2022zdi}, preliminary ideas for the extension of the local analytic subtraction method have been presented in Ref.~\cite{Magnea:2022twu,Magnea:2024jqg}, and there has been extensive recent activity in computing the necessary limits~\cite{Catani:2021kcy,Czakon:2022fqi,DelDuca:2022noh,Catani:2022hkb,Catani:2022sgr,Braun-White:2022rtg,Czakon:2022dwk,Dhani:2023uxu,Craft:2023aew,Cieri:2024ytf}. 

One thing that is important to remember is that the impact of EW corrections can be as sizeable as, or larger than, some of these higher-order QCD effects. In particular, it will be important to determine the EW corrections for ggF Higgs boson production at high $p_T$, where they may be sizeable~\cite{Becchetti:2018xsk,Bonetti:2020hqh,Becchetti:2021axs,Bonetti:2022lrk,Davies:2023npk}.
In particular, negative corrections are expected to reach about ten percent in the high-energy limit where Sudakov logarithms become dominant, as customary at the LHC.
In that respect, NLO EW corrections for di-Higgs production via gluon-fusion have recently been computed~\cite{Davies:2023npk,Bi:2023bnq} and found to be at the level of $10\%$ for large Higgs-boson transverse momentum.

There is a caveat regarding higher order QCD calculations involving jets in the final state. The most common jet radius for the experiments to use is $R=0.4$. This has been shown to result, in some cases, from accidental cancellations that reduce the apparent size of the theoretical uncertainty~\cite{Dasgupta:2016bnd,Rauch:2017cfu,Bellm:2019yyh, Buckley:2021gfw}. This is currently not completely understood.

The advent of N3LO-accurate partonic cross sections has intensified the demand for parton distribution functions (PDFs) at the same order, in order to achieve full N3LO accuracy for hadronic observables.
There has been progress in fitting approximate PDFs at this order, both from the MHST~\cite{McGowan:2022nag} and the NNPDF collaborations~\cite{NNPDF:2024nan}, and especially regarding the evolution (see the one-pager on benchmarking N3LO evolution in this proceeding). 
Unfortunately, producing PDFs at N3LO accuracy is hampered by the lack of calculations for various input processes at this order. This, in turn, makes it difficult to estimate the uncertainty due to missing higher order corrections when producing approximate N3LO PDFs. Of course, one expects that, as calculations improve, more input processes will become available, opening the possibility to have full N3LO PDFs. We emphasize that progress in this direction is essential in order to obtain N3LO accuracy in hadronic predictions.
 

In some cases, the calculations outlined in this one-pager are total cross sections; in other cases differential distributions can be produced. The differential distributions may or may not be amenable to incorporation into parton Monte Carlo predictions, but they remain the highest precision predictions available, and note should be taken of them in any experimental measurements. It is worthwhile to emphasize that the addition of a parton shower should not have a notable impact on any prediction for a reasonably inclusive cross section, such as the Higgs boson transverse momentum~\cite{Bellm:2019yyh}. If such a deviation is found, then it is usually indicative of a mismatch in the calculation. Non-perturbative effects should also vanish at transverse momenta significantly above threshold. 

Many measurements at the LHC have errors that are statistically limited. This will change once the full data-set of the HL-LHC has been taken, and systematic errors may dominate. A conservative estimate is to assume that the systematic errors stay constant. However, this is too pessimistic and it is probable that the systematic errors will also be reduced due to improvements in calibration and analysis, making improvements in theoretical precision even more necessary~\cite{Belvedere:2024wzg}.  Estimates of the current/expected experimental uncertainties for the processes in the wishlist have been provided in previous iterations of the wishlist (see Ref.~\cite{Huss:2022ful}), and will be updated again in the Les Houches 2023 version. 

%% file: sm_pheno/N3LO_benchmark/N3LO_benchmark_main.tex
\section{N$^3$LO PDF Evolution Benchmarking}
\emph{\noindent A.~Cooper-Sarkar, T.~Cridge, F.~Giuli, L.~Harland-Lang, F.~Hekhorn, J.~Huston, 
G.~Magni, S.~Moch, R.S.~Thorne}
\vspace*{0.5em}                                 
 
\noindent For about 20 years the splitting functions for PDF evolution have been 
known~\cite{Moch:2004pa,Vogt:2004mw} at
${\cal O}(\alpha_S^3)$, i.e.\ NNLO in perturbative QCD, and NNLO PDFs have 
become standard in the field. Of course, since the PDFs are produced by 
various groups using different evolution codes it is necessary to check that 
there is indeed consistency between the evolution performed by the various 
groups. To this end a set of benchmark tables were produced in 
\cite{Dittmar:2005ed} in order to provide a means of checking the accuracy of 
any other NNLO evolution code, and indeed this has led to the corrections of 
minor bugs in initial versions of various evolution codes.

Over the past few years there has been an enormous improvement in the amount of 
information available about the ${\cal O}(\alpha_S^4)$, i.e.\ N$^3$LO splitting 
functions. There is numerous results on the low Mellin 
moments~\cite{Moch:2017uml,Moch:2021qrk,Falcioni:2023luc,Falcioni:2023vqq,Moch:2023tdj},
which together with supplementary details from soft gluon resummation in the 
threshold region $x\to 1
$~\cite{Davies:2016jie,Moch:2017uml, Henn:2019swt,Duhr:2022cob,Dokshitzer:2005bf,Almasy:2010wn} 
provides a very good constraint on the higher $x$ splitting functions, particularly 
in the nonsinglet case, and increasingly for the singlet case. There are 
also exact expressions in the limit of large number of 
flavors~\cite{Davies:2016jie,Moch:2017uml,Falcioni:2023tzp,Gehrmann:2023cqm}. 
Leading, and in one case the subleading 
logarithms of the form $\ln^n(1/x)/x$ have also been calculated 
in \cite{Fadin:1975cb,Kuraev:1976ge,Lipatov:1976zz,Kuraev:1977fs,Fadin:1998py,Jaroszewicz:1982gr,Ciafaloni:1998gs,Catani:1994sq} and more recently~\cite{Davies:2022ofz} subdominant logs of the form $\ln^n(1/x)$ were made available. 

Correct evolution including changes of parton number across heavy flavour 
transition points at N$^3$LO also requires knowledge of the 
${\cal O}(\alpha_S^3)$ transition matrix elements. Again, there have been huge 
strides in the 
calculation of these \cite{Kawamura:2012cr,Bierenbaum:2009mv,Ablinger:2014vwa,Ablinger:2014nga,Blumlein:2021enk,Ablinger:2014uka,Ablinger:2014tla,ablinger:agq}.  
Indeed, extremely recently the complete expressions 
for the final matrix elements $A_{gg,H}$ and $A_{Hg}$ were 
completed \cite{Ablinger:2022wbb,Ablinger:2024xtt}.  

The first PDFs using approximate N$^3$LO evolution were obtained by 
MSHT~\cite{McGowan:2022nag}, using a subset of the currently available information. 
The unknown detail in the functions was parameterised in terms of the 
most leading unknown small $x$ term multiplied by a nuisance parameter which 
had both prior and posterior (to a global fit) central values and  
uncertainties.  Very recently NNPDF have used the up-to-date information of 
splitting functions to obtain aN$^3$LO PDFs \cite{NNPDF:2024nan}. The authors of recent 
developments in the calculation of the splitting functions have also produced 
updated estimates of the central values and corresponding uncertainties 
FHMRUVV \cite{Falcioni:2023luc,Falcioni:2023vqq,  Moch:2023tdj}. 

Therefore, given the variety of somewhat differing approximations to full N$^3$LO
splitting functions, and hence evolution, it seems appropriate to repeat the 
type of benchmark comparisons in \cite{Dittmar:2005ed}. In order to maintain 
consistency, and first verify agreement with the code used in \cite{Dittmar:2005ed} 
and by MSHT and NNPDF at NNLO, we maintain the same input conditions as used
in this previous study. In particular we choose the same input PDF combinations and 
parameterizations, and begin evolution at $\mu_{f,0}^2=2~\GeV^2$, which is also chosen 
as $m_c^2$, and define $\alpha_S(2~\GeV^2) =0.35$. We then either evolve using a fixed 
flavour number of $N_f=4$ or a variable flavour number with $N_f=3,4,5$, and compare 
the resulting PDFs at $\mu_f^2 = 10^4~\GeV^2$. Noting some typos in the tables 
in \cite{Dittmar:2005ed}, we do observe good code agreement at NNLO. 

At N$^3$LO, since there are different approximations, we expect some disagreement,which we 
find, although it is only very significant at 
very small $x$. The older MSHT results
diverge to some degree from the more up to date NNPDF and  FHMRUVV results, with 
larger deviation from N$^3$LO at small $x$. However, these regions essentially lie outside those
with significant data constraints, and moreover, 
it has been verified that 
updated benchmark MSHT PDFs, incorporating the new information, agree well in all regions. 
MSHT and NNPDF results from PDF fits 
at approximate N$^3$LO also differ and there are indications from splitting 
function updates that part of this is due to changes in the splitting function input 
used, but part is also due to fit procedures.     
Indeed, we are currently only part of the way towards final complete N$^3$LO PDFs, 
now being in the process of establishing the evolution at this order, with well-controlled, 
and small remaining theoretical uncertainty. 
However, the full PDF determination still lacks a large number of the cross section 
calculations at the necessary order, which are currently incorporated by 
some using some procedure for estimating
missing higher order uncertainties. Progress is also underway in the full calculations 
\cite{Caola:2022ayt,Duhr:2020sdp,Duhr:2020seh,duhr:DY2021,Gehrmann:DYN3LO} 
and will N$^3$LO hadronic cross sections will be systematically incorporated into PDF fits when 
they become available. 

%% file: sm_pheno/N3LOggH/N3LOggH_main.tex
\section{Impact of approximate N3LO PDFs on gluon fusion Higgs production uncertainties}
\emph{\noindent T. Cridge, S. Forte, A. Huss, J. Huston, S. Jones, G. Magni}
\vspace*{0.5em}                                                    

\noindent The measurement of the properties of the Higgs boson at the LHC in Run 3 and beyond at the upcoming HL-LHC and future high energy colliders is a key goal of experimental particle physics in the coming years and decades. A crucial part of this is the investigation of Higgs production and decay processes \cite{Jones:2023uzh,Spira:2016ztx}. For the former the dominant production channel is gluon-fusion producing a Higgs through a quark loop \cite{LHCHiggsCrossSectionWorkingGroup:2016ypw,CMS:2022dwd, Karlberg:2024zxx}. With experimental uncertainties set to be reduced substantially to the 1-3\% level \cite{Cepeda:2019klc,Huss:2022ful}, corresponding progress is required for the theoretical predictions in order to maximise our understanding of the Higgs boson. There are several sources of theoretical uncertainty that contribute \cite{LHCHiggsCrossSectionWorkingGroup:2016ypw,Dulat:2018rbf,Huss:2022ful}; however with recent progress in determining electroweak \cite{Becchetti:2020wof} and finite top mass corrections \cite{Czakon:2021yub,Czakon:2023kqm}, the dominant sources of theoretical uncertainty for inclusive gluon fusion Higgs production are those associated with the PDFs and the value of the strong coupling $\alpha_s$\cite{Heinrich:2020ybq,Huss:2022ful,Jones:2023uzh}.  In this summary, we focus on this aspect.

Efforts within the global PDF fitting groups to improve the precision and accuracy of the PDFs on the experimental, methodological and theoretical sides are continuous and ongoing. In turn this has consequences for the PDF uncertainties associated with gluon-fusion Higgs production, which is our focus, as well as a wide variety of other processes not studied here\cite{Caola:2022ayt}. With the total inclusive $gg \rightarrow H$ cross-section calculated at up to N3LO in QCD\cite{Ball:2013bra,Bonvini:2014jma,Bonvini:2016frm,Ahmed:2016otz,Bonvini:2018ixe,Bonvini:2018iwt,Bonvini:2013kba,anastasiou2014higgs,anastasiou2016high,Mistlberger:2018,ggHiggs,Baglio:2022wzu} and even known in an approximation at 4-loops\cite{Das:2020adl}, but with the PDFs only available up to NNLO in QCD\cite{Bailey:2020ooq,Hou:2019efy,NNPDF:2021njg,PDF4LHCWorkingGroup:2022cjn}, the standard has been to assign an additional uncertainty to the production cross-section due to the mismatch of the order of the PDFs used and the order at which the cross-section is evaluated, representing a theoretical uncertainty from missing higher order corrections in the PDFs. This was taken as half the difference between the cross-section evaluated at NNLO with NNLO PDFs or with NLO PDFs, as an estimate based on experience at previous orders \cite{Anastasiou:2016cez,LHCHiggsCrossSectionWorkingGroup:2016ypw,Dulat:2018rbf,Heinrich:2020ybq,Caola:2022ayt,Jones:2023uzh}. This provided an additional $\sim 1 \%$ uncertainty at $\sqrt{s}=13~{\rm TeV}$, on top of the combined PDF+$\alpha_S$ uncertainty of $\sim 3 \%$. 

In 2022 however the MSHT group produced the first PDF determination at approximate N3LO (aN3LO)\cite{McGowan:2022nag}, by combining recent and older known results for the 4-loop splitting functions, 3-loop transition matrix elements, and N3LO DIS coefficient functions\cite{Moch:2017uml,Moch:2021qrk,Davies:2016jie,Henn:2019swt,Duhr:2022cob,Dokshitzer:2005bf,Almasy:2010wn,Fadin:1975cb,Kuraev:1976ge,Lipatov:1976zz,Kuraev:1977fs,Fadin:1998py,Jaroszewicz:1982gr,Ciafaloni:1998gs,Catani:1994sq,Davies:2022ofz,Kawamura:2012cr,Bierenbaum:2009mv,Ablinger:2014vwa,Ablinger:2014nga,Blumlein:2021enk,Ablinger:2014uka,Ablinger:2014tla,ablinger:agq,Catani:1990eg,Laenen:1998kp,Vermaseren:2005qc}. This constitutes a substantial amount of information about the N3LO PDFs, with remaining missing theoretical ingredients parameterised using theoretical nuisance parameters, whose variation then propagates the uncertainty due to these missing theoretical pieces at N3LO directly to the PDF uncertainties. The result is an aN3LO PDF fit, incorporating the dominant sources of N3LO information into the PDFs, with theoretical uncertainties for unknown elements. This was a substantial step forward in PDF determination, and has very recently been accompanied by an alternative determination of the NNPDF aN3LO PDFs\cite{NNPDF:2024nan}, which include also additional progress made in the past 18 months on the splitting functions and transition matrix elements into the PDFs ~\cite{Falcioni:2023luc,Falcioni:2023vqq,Falcioni:2023tzp,Moch:2023tdj,Ablinger:2022wbb,Gehrmann:2023cqm,Ablinger:2023ahe}.  There has been even further recent progress \cite{Ablinger:2024xtt,Falcioni:2024xyt} which has yet to be incorporated into a PDF fit. In addition, in the NNPDF approach uncertainties related to missing higher order terms can be included through scale variation. This provides a somewhat different methodology for estimating remaining missing higher order corrections beyond N3LO~\cite{NNPDF:2024dpb}.

This progress in PDF determination has the potential to impact both the central value and the assigned uncertainties of gluon fusion Higgs production. Indeed, it was noted by MSHT \cite{McGowan:2022nag} that the aN3LO gluon PDF luminosity around the Higgs mass is smaller than that predicted using an NNLO gluon PDF. This in turn causes the N3LO cross-section + aN3LO PDF result to be smaller than the N3LO cross-section + NNLO PDF prediction. The increase in the total cross section going from NNLO to N3LO is therefore smaller when aN3LO PDFs are used instead of NNLO PDFs, as the increase in the matrix element is partially compensated by the decrease in the PDF. Similar results, albeit reduced in magnitude, have also recently been noted by NNPDF\cite{NNPDF:2024nan}. Our primary focus in this study however is on the uncertainties of $gg \rightarrow H$ production associated with the PDFs. The inclusion of the large amounts of known N3LO information into the PDFs, as well as the incorporation of the theoretical uncertainty for the remaining few missing pieces of information at this order directly into the PDF uncertainties themselves, necessitates a different prescription for the determination of the PDF-associated $gg \rightarrow H$ uncertainties. As a result the separate ``PDF'' and ``PDF-th'' categories utilised previously, are now replaced with a single ``PDF+PDF-th'' category (to which the $\alpha_S$ uncertainty could later be added as before). Utilising the Hessian eigenvectors or Monte Carlo replicas of the MSHT and NNPDF aN3LO PDFs themselves naturally produces this combined uncertainty without the need for ad-hoc prescriptions. The aim of this study is therefore to compare the total PDF uncertainties for gluon fusion Higgs production of the NNLO PDFs with those of the aN3LO PDFs.

We observe that the combined ``PDF+PDF-th'' uncertainty associated with the aN3LO PDFs is smaller than the sum of the two previous `PDF'' and ``PDF-th'' categories computed for the NNLO PDFs, as one would hope with going to higher orders in the PDFs, and despite the larger than anticipated changes in the central values. Both sets see reductions in the uncertainty of $\lesssim 1\%$. This has so far been computed for the public MSHT20aN3LO PDF set\cite{McGowan:2022nag}, and a preliminary version of the NNPDFaN3LO sets, with the public version\cite{NNPDF:2024nan} now available this will be recalculated. More progress is to be expected in these comparisons, whilst on the PDF side the global fitting groups are continuing to update the data, methodology and theory \cite{Cridge:2023ryv,Cridge:2023ozx,Cridge:2024exf} so further comparisons can be made.

An additional important outcome of this study was establishing a framework, based on the recommendations of the Higgs Work Group, in which the impact of new aN3LO PDF sets on the Higgs total cross-section could be assessed and compared. This in turn gives an important input to future recommendations, firstly in addressing whether the previously assigned PDF theory uncertainty was sufficiently conservative, and secondly, in understanding to what extent the different available aN3LO PDF sets give compatible predictions for the Higgs total cross section and the impacts on the total uncertainty

%% file: sm_pheno/H2J/H2J_main.tex
\section{A Critical Study of the gluon-gluon fusion Higgs+2 Jet Background to Vector-boson fusion Higgs Boson Production}
\emph{\noindent X.~Chen, S.~Ferrario Ravasio, Y. Haddad, S.~H{\"o}che, J.~Huston, T.~Je{\v z}o, S.~Pl\"atzer,  C.~T.~Preuss, A.~Tarek, J.~Winter}
\vspace*{0.5em}

\noindent One of the key Higgs boson production processes is through vector boson fusion (VBF). Although sub-dominant, compared to the gluon-gluon-fusion (ggF) production process, the forward kinematics allows for separation of VBF production from the ggF background. (Note that this separation becomes somewhat less efficient at very high transverse momentum~\cite{Buckley:2021gfw}.  The increased statistical power in Run 3 and beyond will allow more precise determinations of the VBF Higgs boson cross section, and thus the Higgs couplings to vector bosons. 

Even with kinematic cuts, there still is a sizeable background from ggF Higgs boson production, and indeed this is one of the largest backgrounds for VBF production, as observed by ATLAS and CMS. The backgrounds from ggF are estimated using parton shower Monte Carlo samples, such as with \powheg+\pythia/\herwig.
The differences between the use of \pythia and \herwig for the parton showering can cause sizeable uncertainties for the level of background; these  uncertainties arise from both the perturbative and non-perturbative aspects of the showers, and deserve to be better understood. As a followup to Ref.~\cite{Buckley:2021gfw}, a study was started at Les Houches in 2023 comparing predictions for Higgs + $\ge$  2 jet production through gluon-gluon fusion from fixed-order calculations to  parton shower Monte Carlo predictions from \sherpa, \herwig, \powheg+\pythia and \powheg+\herwig; predictions for the Monte Carlo programs are at both the parton shower and the hadron levels to test the impact of non-perturbative effects. The non-perturbative tunes include both the recommended tunes from the Monte Carlo authors, as well as the tunes used by ATLAS and CMS. Comparison are being made to STXS cross sections, differential cross sections and a cutflow analysis similar to those used by ATLAS and CMS for the extraction of the VBF signal from the background. Comparisons of a framework that has a variable accuracy as a function of jet multiplicity (\powheg+\pythia/\herwig) to frameworks with a constant (NLO) accuracy (\herwig/\sherpa) are made. A Rivet routine will be made available so that ATLAS and CMS can directly compare to the results of this study. 

The full results will be reported in a separate publication, and linked to this document. 

%% file: sm_pheno/VBFVBS/vbfvbs_main.tex
\section{Common cross section definitions for vector-boson fusion and scattering processes}
\emph{\noindent P.~Azzurri and M.~Pellen summarizing community discussions and efforts.}
\vspace*{0.5em}

\noindent Measurements of vector-boson fusion (VBF) and scattering (VBS) processes at the LHC is of great interest as it offers the possibility to study the self-interactions of gauge bosons in the Standard Model (SM) and explore many associated new-physics scenarios.
The VBF and VBS processes contribute respectively to Vqq and VVqq final states at the LHC, where qq denotes the two final state
quarks (which are associated to the jets) originating from the initial state quarks from which interacting vector bosons are radiated off.
It is worth pointing out that other electroweak (EW) processes, possibly without gauge self-interactions, also contribute to qq$\rightarrow$qqV(V),
for example di-boson or tri-boson processes with one hadronically decaying gauge boson.
Therefore the LHC measurements typically refer to complete EW V(V)qq productions, with
some final state phase-space definition.

The main background sources for EW V(V)qq  measurements are from V(V)jj final state productions (where $\textrm{j} = q \textrm{ or g}$)
with strong interactions, denoted as QCD V(V)jj productions.
Part of these background sources, QCD V(V)qq productions, can also interfere with the signal EW V(V)qq processes,
and additional care is needed to deal with this complication.

An essential specification of the EW signal definition includes a selection of the invariant mass of the two jets ($m_{\rm jj}$),
in order to separate VBF (VBS) productions from VH-type or diboson (triboson) productions where one of the gauge bosons decays hadronically.

Since the early measurements of VBF processes with LHC Run 1 data, and with further measurements of VBF and VBS with Run 2 data,  
signal EW V(V)qq definitions adopted by CMS and ATLAS have been very disparate, also changing with
new measurements of each collaboration.
As a consequence published results are difficult to compare and impossible to combine, calling for the need of an agreement between
experimental collaborations and the theory community to define common cross section definitions to pursue new LHC measurements.
Some initial efforts within the LHC EW working group, and in the context of a multiboson yellow report lead to revive the activity at the  2023  "Physics at TeV Colliders" Workshop.
The discussion in Les Houches took place with  representatives of ATLAS, CMS and of the theory community, resulting in a broad list of directions and recommendations for future VBF and VBS measurements at the LHC, that are listed below.

\begin{enumerate}
\item Define and measure cross sections for  
\begin{enumerate}
    \item full QCD plus EW V(V)qq production with interference,
    \item pure EW  V(V)qq productions without QCD interference, and 
    \item pure EW  V(V)qq productions without $s$-channel (multiboson) contributions. 
\end{enumerate}
\item Define and measure cross sections for both 
\begin{enumerate}
\item parton-level and
\item particle-level signal definitions.
\end{enumerate}
  Priority to particle-level definitions, that can be compared to any prediction with a parton shower, is given.
  Parton level signal measurements can be compared to fixed-order predictions.
\item  Suggestions for particle level  cross sections, with similar kinematic criteria for parton-level definitions.
  \begin{itemize}
  \item Dressed leptons ($\Delta R<0.1$), with angular acceptance $|\eta^\ell|<2.5$.\\
    Transverse momenta $\pT^\ell>20$~GeV for leading and subleading leptons,
    and  $\pT^\ell>5$~GeV for eventual additional leptons (fully leptonic WZ and ZZ VBS). \\
    Invariant mass for same-flavour opposite-sign lepton ($m_{\ell^+\ell^-}>20$~GeV) in order to avoid singular configuration of $\gamma^*\to\ell^+\ell^-$.
     \item Isolated photons $|\eta^\gamma|<2.5$ and  $\pT^\gamma>20$~GeV.
  \item  Anti-$k_{\rm T}$ jets $R=0.4$   with $|\eta^{\rm jet}|<4.7$ and  $\pT^{\rm jet}>30$~GeV.
  \item Angular separation between leptons, photons and jets
    $\Delta R ({\rm jet},\ell)({\rm jet},\gamma) (\ell,\gamma)>0.4 $, or also less stringent on $\ell\gamma$ separation.
    \item Optional requirement on transverse momentum unbalance $\pT^{\rm miss}$ for final states with neutrinos.
  \end{itemize}
\item Define and measure cross sections in a wide array of invariant masses of the VBF/VBS tagging jets (quarks). A suggested  array is : $m_{\rm jj/qq} > 120, 250, 500, 1000, 2000, 5000$ ~GeV.
The more inclusive lower $m_{\rm jj/qq}$ values are of interest also to study the $s$-channel contributions. In the case of processes with hadronic decays of weak bosons, define the tagging jets either as the \pT-leading pair or the pair that maximises $m_{\rm jj}$ 
\item Suggestions for specific differential cross sections:
 \begin{itemize}
  \item Vector-boson transverse momenta : 
  $\pT (\ell, \ell\ell, \gamma)>0, 50, 200, 500 $~GeV, of particular interest for possible pure beyond-the-Standard-Model (BSM) contributions.
\item Azimuthal angular opening of the tagging jets $\Delta\phi_{\rm jj}=[-2.5, -2, -1, 0, 1, 2, 2.5]$,
of particular interest for possible SM-BSM interference contributions.
\item Rapidity opening of the tagging jets,
$| \Delta\eta_{\rm jj}|=[0,2.5,5]$, for possible BSM contributions and also to study QCD-EW interference effects.
 \end{itemize}
\end{enumerate}

Hopefully, recommendations of this type will be adopted for new LHC results, possibly after being further discussed and agreed also in the LHC EW WG.
We would like to emphasis that while the discussion have been focusing on VBF and VBS processes,
all the elements are here provided to extend it to all multiboson processes.

%% file: sm_pheno/VBF_hxswg/VBF_hxswg_main.tex
\section{LHCHXSWG study for vector-boson fusion at the LHC}
\emph{\noindent S.~Ferrario Ravasio and M.~Pellen summarizing community discussions and efforts.}
\vspace*{0.5em}

\noindent Studying in details the properties of the Higgs boson is one of the endeavour of the high-luminosity phase of the LHC.
To that end, precise and appropriate predictions should be provided in order to make the best out of the experimental data.
In that respect, the production of Higgs bosons via vector-boson fusion (VBF) is the second largest production mechanism and therefore deserve particular attention.

This short contribution describe a community study that is taking place within the LHC Higgs Cross Section Working Group (LHCHXSWG) and which benefited significantly of the last Les Houches workshop.
In the following, the expected content of the study is highlighted, results will be made public in a subsequent publication.
The work aims at three different aspects: a review of recent theoretical and experimental work, state-of-the-art predictions at fixed order, and recommendations for parton-shower uncertainties.

\begin{enumerate}
 \item \underline{Short review} \\
 On the one hand, recent theoretical progress for VBF at the LHC will be reviewed.
 On the other hand, the latest experimental findings will be summarised.
 The former is particularly important as it has be realised that theoretical works are not always properly acknowledged in experimental publications.
 We hope to mitigate this problem by providing explicit recommendations for referencing various theoretical aspects of VBF at the LHC.
\item \underline{State-of-the-art predictions at fixed order} \\
 Typically, the predictions that are provided within the LHCHXSWG are inclusive predictions~\cite{LHCHiggsCrossSectionWorkingGroup:2016ypw}.
 While these make perfectly sense for the gluon-gluon--fusion mechanism, 
 they are not so appropriate for VBF which is typically measured in exclusive phase-space regions.
 
 To that end, the fixed-order predictions will be provided in a typical fiducial region for two-dimensional differential distribution in the invariant of the two jets ($m_{\rm jj}$), the transverse momentum of the Higgs boson ($p_{\rm T, H}$), the azimuthal angle between the two jets ($\Delta \phi_{\rm jj}$), and the rapidity difference between the two jets ($\Delta y_{\rm jj}$).
 In addition, predictions will be provided for the STXS bins~\cite{Berger:2019wnu} which are fully inclusive.
 The predictions aims at providing principally NNLO QCD + NLO EW predictions.
 Non-factorisable and loop-induced interferences are also expected to be included.
 \item \underline{Parton-shower corrections and uncertainties} \\
 In the last few years, several works uncovered particular feature of parton-shower corrections in VBF topologies \cite{Ballestrero:2018anz,Jager:2020hkz,Hoche:2021mkv}. 
 Eventually, it has been found that uncertainties related to the perturbative part of parton showers do not exceed about ten per cent~\cite{Hoche:2021mkv,Buckley:2021gfw}.
 
 Using the same phase spaces as for the fixed-order study and the same binnings, several parton-shower predictions will be provided. It is expected that the findings of Refs.~\cite{Hoche:2021mkv,Buckley:2021gfw} will be confirmed.
\end{enumerate}

Note that several other important aspects of VBF at the LHC have been discussed in Les Houches such as 
gluon-gluon--fusion contamination in VBF phase-space, boosted-Higgs topologies, and the potential role of parton-shower tuning (regarding for example MPI, hadronisation, underlying event) and parton-shower uncertainties.
These will be addressed in the future in different publications.

%% file: sm_pheno/EW/EW_main.tex
\section{Theoretical uncertainties for electroweak corrections}
\emph{\noindent A.~Huss and M.~Pellen summarizing community discussions and efforts.}
\vspace*{0.5em}

\noindent Given the current and upcoming precision of LHC measurements, the definition of uncertainties in electroweak (EW) corrections is becoming a particularly pressing issue.
While for QCD corrections, simple prescriptions to estimate missing higher-order corrections exist (renormalisation- and factorisation-scale variations), it is not the case for the EW case.
It would therefore be particularly useful to set-up some common prescriptions for estimating uncertainties related to the calculation of EW effects.
While a one-size-fits-all procedure is likely unfeasible, a set of prescriptions and guidelines to estimate various aspects of theory uncertainties associated with EW corrections can be defined. 

The understanding of EW corrections is by now well advanced~\cite{Denner:2019vbn,Andersen:2014efa} and an attempt in estimating missing higher-order corrections have been made in the context of Standard Model background for dark-matter searches~\cite{Lindert:2017olm}.
The main hurdle in defining a prescription stems from the fact that EW corrections can have very different origins of enhancements.
In a discussion in Les Houches, several sources with corresponding possible prescription have been identified.

\begin{description}
\item[Renormalisation part:]
One part of the EW corrections is related to the renormalisation of UV divergences.
In this context, comparing different renormalisation schemes can be useful also in view of assessing potential missing higher-order terms.
Nonetheless, it should be kept in mind that in a particular context, some schemes are more well-motivated than others and therefore a naive variation of schemes can over-estimate true uncertainties and such a prescription needs to be implemented with care.
\item[Sudakov logarithms:]
At the LHC, various processes are increasingly probed deeper in the high energy limits.
In this regime, so-called Sudakov logarithms can become enhanced through the ratio of the gauge-boson mass and the characteristic scale of the process appearing in the arguments.
On the other hand, they are known to factorize and to be universal~\cite{Denner:2000jv}, such that their impact at each order in perturbation theory can be assessed using prescriptions based on those properties.
In addition, there has been a revived interest in providing codes for computing these contributions~\cite{Bothmann:2020sxm,Pagani:2021vyk,Lindert:2023fcu} making these type of corrections more accessible also to non-experts.
\item[QED final-state radiation:]
With the availability of several public tools to compute these corrections, it is anticipated that prescriptions can be formulated to assess the size of missing higher-order (and higher-logarithmic) corrections. 
\item[$\gamma$-induced contributions:]
Such contributions are typically suppressed and have a small uncertainty since the adoption of the LuxQED method~\cite{Manohar:2016nzj}.
As for any other parton distribution function (PDF), its uncertainty can be obtained by varying the replicas provided by the PDF sets.
Constituting separate partonic channels, an additive approach to include these contribution to the prediction is well motivated.
\item[Heavy gauge-boson radiation:]
Such real-emission contributions are known to partially compensate the (virtual) Sudakov logarithms. 
While their computation is in principle rather straightforward, their inclusion depends strongly on the details of the experimental analysis, e.g.\ if these sub-processes are considered as background, etc.
As such, defining a common prescription appears more challenging in this case and a case-by-case treatment appears to be necessary.
\end{description}

Another non-trivial aspect relates to the fact that these various EW effects are enhanced in different regions of phase space.
Consequently, some interpolation will likely be necessary.

In addition to purely EW corrections, mixed QCD--EW corrections also appear at next-to-next-to-leading-order accuracy.
It will therefore also be desirable to obtain a prescription to estimate the related uncertainties.

This summary only provides a snapshot of the past and on-going discussions on this topic.
We hope to be able to reach a community prescription and agreement that will be documented in a separate note.

%% file: sm_pheno/flavouredjets/flavouredjets_main.tex
\section{Infra-Red and Collinear Safe definitions of jet flavour}
\vspace*{0.5em}

\noindent Hadronic jets containing heavy flavours, i.e.\ charm and beauty, are central to the LHC physics programme: they are important for Higgs and top studies, determination of parton distribution functions and new physics searches. At the LHC, flavoured jets are usually identified by the presence of $B$ or $D$ hadrons inside already reconstructed anti-$k_t$ jets: experimental analyses exploit displaced vertices inside the detector or fully reconstruct the complete decay chain of the heavy hadrons. 
 This has led to the standard approach of defining the flavour of an anti-$k_t$ jet, in simulation and predictions, according to the presence of heavy hadrons associated with it. 
From a theoretical point of view, such a definition needs to be revised, as it is infrared unsafe when performing perturbative calculations, leading to a sensitivity to the soft and collinear region~\cite{Banfi:2006hf}. In particular, this leads to uncancelled divergences at two loops and beyond when adopting calculation schemes in which the mass of the heavy flavour is not retained, as often assumed in state-of-the-art theoretical predictions. 

Recently, several flavour-dependent jet algorithms have been proposed~\cite{Caletti:2022hnc,Czakon:2022wam,Gauld:2022lem,Caola:2023wpj}, designed to overcome infrared safety (or sensitivity) issues by interleaving kinematics and flavour information in the jet clustering. In perturbation theory, these algorithms are infrared safe to all orders (or up to high order), and they feature exact (or close to exact) anti-$k_t$ kinematics. As such, they can be safely employed in theory predictions, and several results for processes featuring flavoured jets at NNLO accuracy have appeared~\cite{Czakon:2022wam,Gauld:2022lem,Czakon:2022khx,Hartanto:2022ypo,Gauld:2023zlv,Gehrmann-DeRidder:2023gdl}. Discussions have occurred in Les Houches regarding the feasibility of an experimental implementation of the theory-friendly flavoured algorithms. The current formulation of the flavoured algorithms relies on the knowledge of all the flavoured particles in the event, which is usually unavailable in an experimental context. In particular, the labelling of jets in the presence of double (or multi) $b/c$ flavours has been identified as a significant bottleneck. While a jet containing a $B$ or $D$ hadron and its corresponding anti-hadron is flavour neutral, its experimental signature is more similar to a $b/c$-jet rather than a gluon jet.

Given the intricate mismatch between the experimental and theory approaches in assigning flavour to jets, a reliable data-to-theory comparison is not possible, and delicate Monte Carlo correction steps are required to bring theory predictions closer to the (infrared-sensitive) flavour assignment strategies used by the experimental collaborations. Crucial and so far unexplored aspects are the accuracy of these corrections and reliable estimates of the associated uncertainties. Further, it is valuable to understand how the choice of flavoured jet algorithm influences these properties. The algorithms’ computational performance is another feature vital for the implementation in experimental analysis and, therefore, needs to be studied and optimised.

To this end, a detailed and consistent comparison of the flavoured jet algorithms started in Les Houches and is currently ongoing. As the first step, a common framework with a Fastjet~\cite{Cacciari:2011ma} implementation of four flavoured jet algorithms was created and made public (\url{https://github.com/jetflav}), a publication as a FastJet contribution \href{https://fastjet.hepforge.org/contrib/}{fjcontrib} is anticipated. Secondly, on a purely phenomenological level, the algorithms are being compared both in fixed-order calculations, up to NNLO, and in LO+PS or NLO+PS simulations. A variety of benchmark processes (such as vector boson plus jet production ($Z+b/c$, $W+c$), associated production of $W$ and Higgs boson with a subsequent $H \to b\bar{b}$ decay, or pure QCD scattering) are being investigated in several scenarios, with fiducial cuts either in the central or forward rapidity region. The plan is to focus on standard observables, e.g.\ the transverse momentum of the flavoured jet, with the possibility of including jet substructure observables, e.g.\ the jet mass in the analysis.
Such comparisons are useful for understanding the differences between the various algorithms once applied to physical observables. Moreover, they will be used to shed light on the MC corrections needed for theory-data comparisons. The uncertainty of these corrections can be estimated by varying matching schemes and PS providers, with and without multiple parton interactions and hadronisation effects. The group aims to prepare a document containing the results of these comparisons that will serve as a guideline or recommendation for employing the new flavour-sensitive algorithms for LHC phenomenology.

%% file: sm_pheno/ttH_STXS/ttH_STXS.tex
\section{Studies on possible STXS binning options for $t\bar{t}H$ production}
\vspace{0.5em}

In the SM, the Higgs boson is predicted to couple most strongly to the top quark. The top-quark Yukawa coupling, expected to be of order unity, can be probed directly by measuring the $t\bar{t}H$ cross-section. With the full LHC Run 2 dataset the precision of differential measurements obtainable has increased. These differential
cross sections can be extracted using the STXS~\cite{Andersen:2016qtm,Berger:2019wnu}
(Stage 1.2~\cite{Amoroso:2020lgh}) formalism, discussed and defined at previous Les Houches workshops. For $t\bar{t}H$ the current recommendation utilises a binning in the transverse momentum of the Higgs, $p_{\mathrm{T}}(H)$. 

Measurement of $t\bar{t}H$ production is a key part of testing the Higgs mechanism, understanding
the quantum stability of the Universe through the interplay between top Yukawa coupling and Higgs self
couplings~\cite{Hambye:1996wb,PhysRevD.44.3620,Casas:1994qy,Isidori:2001bm,Espinosa:2007qp,Buttazzo:2013uya,Degrassi:2016wml,DiVita:2017eyz}, and to test the Higgs sector as a source of CP
violation~\cite{Farrar:1993sp,Bernreuther:2002uj,Brod:2013cka,Ellis:2013yxa,Demartin:2014fia,Khatibi:2014bsa,Boudjema:2015nda,Demartin:2015uha}.  

At the LHC the CP properties of the top quark
 have an impact on the production rates
\cite{Ellis:2013yxa,Boudjema:2015nda,AmorDosSantos:2017ayi} and also in the kinematic
distributions. Measurement of the top-Higgs system can also give sensitivity to the self-coupling beyond the inclusive $t\bar{t}H$
cross-section~\cite{Maltoni:2017ims}.

Given these facts, at Les Houches discussions and preliminary initial studies were started to investigate possible modifications to the current STXS $p_{\mathrm{T}}(H)$ binning and/or the use of different observables. This included discussions on the possible use of variables inspired by Refs.~\cite{Boudjema:2015nda,Ferroglia:2019qjy,Cao:2020hhb} such as angular variables defined in the $t\bar{t}H$ centre-of-mass frame. Studies were also performed to consider possible improvements in sensitivity to the Higgs self-coupling from alternative observables to $p_{\mathrm{T}}(H)$ inspired by Ref.~\cite{Maltoni:2017ims}, for example $m(t\bar{t}H)$. Initial studies showed modest increases in sensitivity for $m(t\bar{t}H)$ but not yet sufficient justification for a new STXS recipe. 

These studies at Les Houches were very preliminary and somewhat limited in scope. More studies are certainly required before any modification to the STXS formalism could be proposed and these studies should continue in the context of the LHC Higgs WG.

%% file: tools/tools.main.tex
\begingroup
\let\clearpage\relax
\subimport{tools/uncertainties/}{uncertainties_main}
\subimport{tools/correlations/}{correlations_main}
\subimport{tools/EIC/}{EIC_main}
\subimport{tools/GPU/}{GPU_main}
\subimport{sm_pheno/negativeweights/}{negativeweights_main}
\subimport{tools/reweighting/}{reweighting_main}
\subimport{tools/unfolding_ML/}{unfolding_ML_main}
\subimport{tools/MC4SM/}{MC4SM_main}

\subimport{tools/WDR/}{WDR_main}

\subimport{tools/interfaces/}{interfaces_main}
\subimport{tools/reproducibility/}{reproducibility}
\endgroup

%% file: tools/uncertainties/uncertainties_main.tex
\section{Defining better uncertainties on the choice of MC model}
\emph{\noindent M.~LeBlanc, J.~McFayden and S.~Plätzer summarizing community discussions and efforts}
\vspace*{0.005em}

\noindent Differences between predictions from models of hadronic radiation continue to be a leading source of systematic uncertainty at the LHC, limiting the precision of numerous Run 2 physics results.
These uncertainties manifest in two main ways.
First, they can come from direct comparisons  between models within a particular search or measurement, \emph{e.g.} by changing from the \textsc{Pythia} to the \textsc{Herwig} parton shower.
Second, they can arise indirectly through other systematic uncertainties such as those related to the calibration of the Jet Energy Scale~\cite{JETM-2022-01,JETM-2018-05,JETM-2018-02} or JSS-based classifiers~\cite{JETM-2020-02,ATL-PHYS-PUB-2023-032,ATL-PHYS-PUB-2023-020,ATL-PHYS-PUB-2022-039,CMS-JME-18-002,CMS-DP-2022-005}, where MC-to-MC comparisons are made during the calibration procedure.
The \emph{ad hoc} nature of these so-called ``Modelling Uncertainties,'' combined with their significant impact, has long been a source of frustration for the LHC community.
At Les Houches, several discussions took place that aimed to summarise the current state of affairs, and identify a path towards better uncertainties on the choice of MC models in the future.

\subsection{Current approaches}

Presently, `monolithic' two-point modelling uncertainties are frequently made.
Such comparisons vary all aspects of generator setups simultaneously, \eg by swapping a nominal MC sample generated with \pythia\ for an alternative generated with \herwig\ that uses a different matrix element calculation, parton shower algorithm, hadronisation model, PDF set, \emph{etc.}
%
Some significant effort has been recently made by the ATLAS Collaboration to improve on such monolithic comparisons in the context of the jet energy scale (JES) uncertainties~\cite{ATL-PHYS-PUB-2022-021,ATL-JETM-2022-005}.
The `jet flavour response' component of the JES uncertainty accounts for differences in the response of jets that are initiated by gluons, rather than quarks, and was previously determined by comparing the gluon-initiated jet response between \pythia and \herwigpp~\cite{JETM-2018-05}.
This comparison previously resulted in a large uncertainty that was the dominant source of uncertainty on the ATLAS JES\footnote{The situation within the CMS Collaboration for this particular source of uncertainty is similar: the same generator comparison is made and also results in a dominant uncertainty across a comparable kinematic region~\cite{CMS-JME-13-004}.} for jets with transverse momenta roughly between 60~GeV and 300~GeV over the course of both Run 1 and Run 2.
In the new approach, multiple two-point comparisons vary specific aspects of generator setups: \emph{e.g.} comparing \herwig\ with angle-ordered and dipole parton shower models, \sherpa\ using default cluster-based or \pythia\ string-based hadronisation models, \emph{etc.} while holding other aspects of the setups (PDFs, etc.) constant wherever possible.
Making more factorised comparisons results in a total flavour response uncertainty that is reduced by roughly a factor of 3 for jets with \pt=60~GeV, resulting in an updated uncertainty component that is now subdominant relative to other sources of uncertainty on the ATLAS JES.

\subsection{Future directions}

The more factorised two-point comparisons studied by ATLAS were recognised as a step forward; however, some guidance was provided by members of the Les Houches community regarding further improvements to this methodology.
In particular, it was pointed out that there can be significant interplay between hadronisation and parton shower algorithms, and that a more reliable way to assess such two-point uncertainties would be to use grids of MC setups that are tuned and configured consistently by the MC authors themselves (\eg a 2-by-2 grid of \herwig setups with different parton shower and hadronisation algorithms).
Such provisions would be welcomed by the community, and they may also present an opportunity to harmonise the tuning strategies and reference datasets used by the different MC authors.

In the medium-term future, the Les Houches community is optimistic about the expected impact of NLL parton shower algorithms in this area.
Comparisons of such algorithms should provide meaningful uncertainties on the underlying perturbative physics.
Non-perturbative models should also be revisited, to avoid mismatched levels of progress of MC generator components.
Improved understanding of the correlations between hadron species within jets and studies of universality between both \ee\ and \pp\ systems, and low- and high-\pt jets were identified as possible avenues for development.

It was suggested that the growing number of differential measurements of JSS observables that have been performed at the LHC with Run 2 data may be an existing place where progress can be made towards improved understanding.
Another potentially promising avenue are studies with archival LEP data, which can be difficult to access for experts outside of the LEP communities.
However, successful recent efforts have been made with these archival data~\cite{Chen:2022qod,Chen:2023rif}: the Les Houches community is hopeful that such publications indicate that new members of the particle physics community are gaining expertise with lepton collider data analysis that will have a positive impact on MC modelling in the years to come.


%% file: tools/correlations/correlations_main.tex
\section{Modelling survey and correlation analyses}
\vspace*{0.5em}

\noindent The interplay between parton showers and hadronization, and the quest for comprehensive uncertainty estimates, becomes a central need for the community. In particular, a deeper understanding of observables which probe the interface between parton showers and hadronization and might be suited to test constraints from shower variations and tuning, is needed. One working group is studying such a list of observables at hadron colliders, and has also developed new observables which focus on leveraging correlations among jet constituents to reveal the built-up of jet structure.

The following Rivet~\cite{Buckley:2010ar} routines were included in the survey:
 \\
 ATLAS\_2020\_I1790256 -- 13 TeV Lund jet plane~\cite{ATLAS:2020bbn} (Dijets, pT $\geq$ 675 GeV), ATLAS\_2020\_I1808726 -- 13 TeV Event Shapes (Thrust etc.)~\cite{ATLAS:2020vup} (Multijets, HT2 $\geq$ 1 TeV), ATLAS\_2019\_I1772062 -- 13 TeV Soft-drop mass, rg, zg~\cite{ATLAS:2019mgf} (Dijets, pT $\geq$ 300 GeV), ATLAS\_2019\_I1724098 (MODE="DJ") -- 13 TeV jet tagging observables~\cite{ATLAS:2019kwg} (Dijets, pT $\geq$ 400 GeV), ATLAS\_2019\_I1740909 -- 13 TeV nTrk, fragmentation functions~\cite{ATLAS:2019rqw} (Dijets, pT $\geq$ 300 GeV), ATLAS\_2018\_I1634970 -- 13 TeV Inclusive Jets~\cite{ATLAS:2017ble} (Inclusive jets, pT $\geq$ 100 GeV), CMS\_2021\_I1920187 [MODE="DIJET"] angularities in Z+jet and multijets~\cite{CMS:2021iwu} (pT $\geq$ 50 GeV binned up to 1 TeV), CMS\_2018\_I1682495 jet mass in dijets~\cite{CMS:2018vzn} (pT $\geq$ 200 GeV), CMS\_2021\_I1972986 13 TeV inclusive jets~\cite{CMS:2021yzl} (pT $\geq$ 97 GeV), CMS\_2021\_I1920187 [MODE="ZJET"] angularities in Z+jet and multijets~\cite{CMS:2021iwu} (pT $\geq$ 50 GeV binned up to 1 TeV), and CMS\_2018\_I1690148 jet substructure observables in ttbar~\cite{CMS:2018ypj} (pT $\geq$ 30).

The community does not have Rivet routines from all measurements that would be useful\footnote{The current coverage of experimental results in Rivet can be found at \url{https://rivet.hepforge.org/rivet-coverage}.}.
Routines are particularly missing for ALICE measurements that would otherwise be extremely valuable~\cite{ALICE:2021bib,ALICE:2021dhb,ALICE:2022phr}, and in particular correlation measurements seem to be promising to further learn about differences and short-comings of present models.
%

%% file: tools/EIC/EIC_main.tex
\section{Photoproduction at the EIC}
\emph{\noindent I.~Helenius, P.~Meinzinger, S.~Plätzer}
\vspace*{0.5em}

\noindent The prospect of the EIC as one of the next colliders also raises questions to what extent multi-purpose event generators are ready to be used for the physics of a state-of-the-art electron-proton collider. For several years, benchmarking against HERA data has been quite difficult due to the lack of analyses usable as Rivet plugins, a situation which now has changed significantly. While DIS, including event shapes and jet production, has been compared to the LHC-age multi-purpose event generators in the realm of developing new parton shower and matching and merging algorithms, photoproduction is one class of reactions where new comparisons are needed in order to thoroughly test end evaluate the uncertainties of existing approaches. During this Les Houches workshop, one group has formed which is comparing \herwig~7, \pythia~8 and \sherpa~3 for photoproduction at LEP, HERA and the EIC. We use existing data for LEP and HERA, and study a class of observables similar to those addressed in \cite{Meinzinger:2023xuf} to compare predictions at typical EIC energies. A detailed paper for this study is in preparation
and can serve as a basis for further development. 

%% file: tools/GPU/GPU_main.tex
\section{Matrix-element generation on GPUs and vector CPUs}
\label{sec:gpu}
\vspace*{0.5em}

\noindent Driven by rapidly evolving computing hardware, various discussions in Les Houches focused on porting code for fixed-order perturbative calculations from CPUs to GPUs. An important aspect is to reach a production level code base and its effective implementation in the experimental tool chains. Work and discussions at Les Houches initiated an effort to better understand the related requirements and triggered follow-ups with ATLAS and CMS after the workshop that are still ongoing. In the case of \mgamc, the discussions and hands-on work together at Les Houches and their later follow-up enabled CMS to perform useful tests, of both the GPU and vectorized CPU implementations, which were presented at a CERN MC workshop in November 2023~\cite{NLOWS2023}, with further updates foreseen for CHEP2024. Similar exchanges with ATLAS at Les Houches were essential for identifying some LO physics processes that are relevant to ATLAS, which later led to a large effort to debug and fix some issues in these processes in \mgamc on GPU. In the case of \chilipepper, the code base was published~\cite{Bothmann:2023gew} and made available to general users in a form that allows production-level performance in the case of leading-order high-multiplicity simulations~\cite{Bothmann:2023ozs}.

Looking further ahead the discussions begun at Les Houches will be continued for algorithms beyond tree-level, where the hand-off between GPU and CPU may trigger problems with floating point precision, especially in light of the anticipated relative decrease in the number of FP64 cores on newer GPUs. This point is also related to a more general discussion regarding preparations for future hardware. In this regard it is also important to understand which HPC allocations can be expected to be obtained by the community in the long term, and how dependable those allocations will be. This topic was again discussed at the CERN workshop~\cite{NLOWS2023} and noted as one of the drivers for decisions on how to design future code bases.

An important related topic is the sharing of resources between different experiments, and between experiment and theory. Experiments typically expend significantly more computing time for event generation on the grid than any theory group has access to. A stronger connection between the different user groups and also between developers and users might be mutually beneficial. For instance, access to grid-scale resources might help for computations currently out of reach, e.g.\ inclusion of $V+4/5$j@NLO~\cite{Berger:2010zx,Ita:2011wn,Bern:2013gka} in particle-level simulations, or fixed order calculations of NNLO or even N3LO predictions~\cite{Prestel:2021vww,Bertone:2022hig}.

A discussion on the benchmarking of existing GPU ports was started at Les Houches. The two teams developing \mgamc and \chilipepper on GPUs are working on very similar strategies to port or re-develop matrix-element generators that may be used at scale~\cite{Valassi:2021ljk,Valassi:2022dkc,Valassi:2023yud,Hageboeck:2023blb,Bothmann:2021nch,Bothmann:2023siu,Bothmann:2023gew}. Different stages of the integration and event generation are implemented differently, however. It will be beneficial to see how they compare and if there are bottlenecks that can be avoided. This would include a detailed profiling of \mgamc and \chilipepper on GPUs and vector CPUs, comparing the computational pros and cons of different approaches like helicity summation versus sampling, Feynman diagrams versus recursion relations and comparing timings and profiles for a few standard candle processes. The importance of a coherent development plan was emphasized in the LHCC report~\cite{CERN-LHCC-2022-007}
on HL-LHC software and computing and in the community papers prepared for it~\cite{HSFPhysicsEventGeneratorWG:2020gxw,Campbell:2022qmc}.

Currently, both \mgamc and \chilipepper only support LO processes in their GPU ports, but plans exist to extend this work to QCD NLO processes~\cite{Wettersten:2023nlo}. The discussions in Les Houches, including experts in fixed-order calculations, were useful to identify potential issues. In the case of \mgamc, some Les Houches discussions about local subtraction schemes, and their later follow-up with NLO matching experts, helped to better clarify a strategy for data parallelism in NLO generation workflows using the FKS subtraction scheme~\cite{Frixione:1995ms}.

%% file: sm_pheno/negativeweights/negativeweights_main.tex
\section{Negative Weight Reduction for the Higgs diphoton background}
\emph{\noindent J.~R.~Andersen, A.~Cueto, S.~P.~Jones, A.~Maier}
\vspace*{0.5em}

\noindent Monte Carlo event generators are a critically important tool for studying physics at particle colliders.
General Purpose Event Generators, such as \herwig~\cite{Bellm:2015jjp}, \sherpa~\cite{Sherpa:2019gpd} and \pythia~\cite{Bierlich:2022pfr}, are used almost ubiquitously to produce theory predictions for both signal and background processes at the Large Hadron Collider (LHC). They are often combined with fixed-order and matching tools such as \madgraph~\cite{Alwall:2014hca} or \powheg~\cite{Alioli:2010xd}.
As we enter the high-luminosity (HL-LHC) era, the computing demands, both in terms of CPU cycles and disk space, required to produce sufficiently precise simulations with these toolchains are considerable and justify the investigation of methods for improving the efficiency of unweighted event generation~\cite{HEPSoftwareFoundation:2017ggl,CERN-LHCC-2022-005,HSFPhysicsEventGeneratorWG:2020gxw}.

At next-to-leading order (NLO), due to presence of subtraction terms in the fixed-order calculation (and potentially the matching and merging to parton showers, the weight of the simulated events is not strictly positive.
The presence of negatively weighted events reduces the statistical power of an event sample and necessitates the production of larger event samples to obtain a given statistical precision.
The impact of the negative events can persist even after unweighting, with typical unweighting procedures yielding events set with a given weight $\pm w$, where $w$ is a positive constant determined by the details of the process and generator.
A simulation that produces unweighted events with both positive and negative signs can attain the same statistical accuracy as one that produces events with only a positive weight by generating a number of events which is larger by a factor
$c(r_-) = 1/(1-2 r_-)^2$, where $r_-$ is the fraction of negatively weighted events~\cite{Frederix:2020trv} (neglecting correlations between the negative and positive events). 
In the last few years, several techniques have been developed to address the problem of negatively weighted events. Broadly, they are based on avoiding negative events during generation (for recent progress see \cite{Olsson:2019wvr,Frederix:2020trv,Gao:2020vdv,Bothmann:2020ywa,Gao:2020zvv,Danziger:2021xvr}), and resampling after generation (see e.g.~\cite{Andersen:2020sjs,Nachman:2020fff,Verheyen:2020bjw,Andersen:2021mvw}). 
In the following we will discuss one such approach called the Cell Resampler~\cite{Andersen:2021mvw,Andersen:2023cku}.

In brief, the cell resampler works by identifying the most negative events in the sample and distributing their weight to \textit{nearby} events such that the number of events in the sample with a negative weight is reduced.
The algorithm used by the cell resampler relies on constructing infrared safe objects from each event (such that "nearby" is interpreted in an IR safe manner) and on a \textit{distance metric} between events that defines how close two events are considered to be.
In this we apply the metric discussed in Ref.~\cite{Andersen:2021mvw,Andersen:2023cku}, but other metrics are desirable, as discussed later.
It is often desirable to prevent the cell resampler from averaging the weight of distant events, so as not to distort observables beyond the resolution required, even if this will leave some negative weights in the sample.
The maximum distance at which two events are allowed to be combined is controlled by the maximum cell size parameter.

The goal of the present study was to apply the so-called cell resampling technique to an event sample similar to that which is used by experimental collaborations for the estimation of background processes.
In particular, we focus on the simulation of the diphoton background relevant for the experimental study of the Higgs boson production and decay process $pp \rightarrow H(\rightarrow \gamma \gamma) + \mathrm{jets}$.
At NLO, this particular background is plagued by a relatively large negative event fraction.

Our study begins by generating a large set of unweighted $p p \rightarrow \gamma \gamma + \mathrm{jets}$ events.
We aim to start with a sample that is as close as possible to that used by the experimental collaborations.
To this end we produce a set of 1M and 10M showered, multi-jet merged and unweighted events using \madgraph interfaced with the \pythia shower. The matrix elements of the calculation were computed at NLO with up to two additional partons. Divergencies in the radiation of photons off quarks was avoided by imposing a Frixione's isolation~\cite{Frixione:1998jh} with parameters $\delta_0=0.1$, $n=2$ and $\epsilon_{\gamma}=0.1$. Jets were reconstructed with the $kt$ algorithm with a jet radius of 1.0 and a minimum jet transverse momentum of 10 GeV. The merging with the parton shower was performed using the FxFx scheme~\cite{Frederix:2012ps} with a merging scale of 20 GeV.
We would like to study the impact of cell resampling on observables that are typical of experimental analyses, we therefore focus on the set of observables produced using a Rivet~\cite{Bierlich:2019rhm} analysis that mimics the experimental cuts previously used by the ATLAS collaboration to study prompt photon-pair production~\cite{ATLAS:2021mbt}.
In the implementation of the cell resampler existing prior to this study, no photon isolation algorithm was present.
For this work, we, therefore, first extend the cell resampler to support fixed cone photon isolation~\cite{Frixione:1998jh}.

\begin{table}[h!]
\centering
\begin{tabularx}{0.8\textwidth}{@{}>{\raggedright\arraybackslash}X|>{\raggedright\arraybackslash}X >{\raggedright\arraybackslash}X|>{\raggedright\arraybackslash}X >{\raggedright\arraybackslash}X@{}}
\hline
Max Cell Size& $r_-^\mathrm{1M}$ & $r_-^\mathrm{10M}$ & $c(r_-^\mathrm{1M})$ & $c(r_-^\mathrm{10M})$  \\ \hline
0                           & 0.297                          & 0.296                           & 6.07         & 6.01   \\
10 GeV                 & 0.296                          & 0.294                           & 6.01         & 5.89   \\
30 GeV                 & 0.287                          & 0.280                           & 5.51        & 5.17   \\
60 GeV                 & 0.261                          & 0.245                           & 4.38         & 3.84   \\
100 GeV                & 0.218                          & 0.195                           & 3.14         & 2.69  \\
$\infty$                 & 0.000                          & 0.000                          & 1               & 1  \\ \hline
\end{tabularx}
\caption{The negative event fractions, $r_-^\mathrm{1M}$ and $r_-^\mathrm{10M}$, and relative costs, $c(r_-)^\mathrm{1M}$ and $c(r_-)^\mathrm{10M}$, obtained after cell resampling on a set of 1M and 10M showered, multi-jet merged and unweighted $\gamma \gamma + \mathrm{jets}$ events.}
\label{tab:cres_yy}
\end{table}

In Table~\ref{tab:cres_yy} we report the negative event fractions and relative costs obtained by running the cell resampler with various maximum cell sizes.
As expected, the larger cell sizes lead to a more significant decrease in the negative event fraction, with an infinite max cell size returning an event set with only positive events.
Furthermore, running on a larger event sample of 10M events, rather than 1M events, gives larger reduction in negative event fraction for a given cell size. 
This is straightforward to understand as the 10M event sample populates the phase-space more densely and so more events can land in each cell for a given max cell size.


\begin{figure}[ht]
  \centering
  \begin{subfigure}[]{
      \includegraphics[width=.45\linewidth] {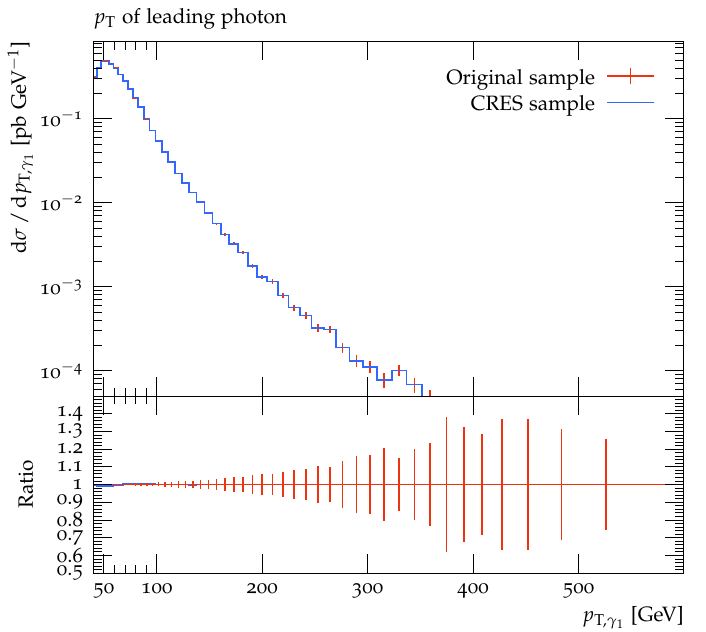}   
      }
  \end{subfigure}
  \hfill
  \begin{subfigure}[]{
      \includegraphics[width=.45\linewidth] {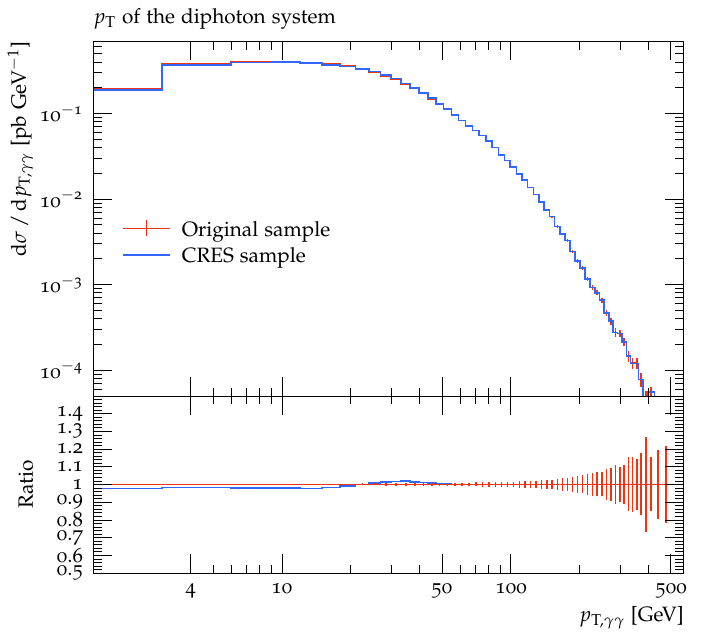}   
      }%
  \end{subfigure}
  \caption{The leading photon $p_T$ and diphoton system $p_T$ obtained using a sample of 10M events before and after cell resampling with a maximum cell size of 30 GeV.
    The error bars indicate the statistical uncertainty on the original Monte Carlo sample.}
  \label{fig:cres_yy}
\end{figure}

In Figure~\ref{fig:cres_yy} we show the leading photon $p_T$ and diphoton system $p_T$ obtained using the larger event sample before and after cell resampling.
We can see that with a maximum cell size of 30 GeV the distributions are largely reproduced faithfully (within the statistical uncertainty).
However, at very small $p_T$ some distortion of the distribution is possible, due to the presence of cells with a size significantly larger than the bin widths. 
This distortion can be cured by reducing the maximum cell size at the cost of a larger negative event fraction remaining in the sample.
Increasing the maximum cell size beyond 30 GeV induces similar distortions at larger values of $p_T$ and increases the magnitude of the distortions.
Furthermore, running on this showered and unweighted sample, the overall negative weight fraction reduction for a given level of distortion (e.g. within the original statistical uncertainty) is not as significant as previously observed in studies of weighted fixed-order computations~\cite{Andersen:2023cku}.

Based on these observations, a follow-up study is ongoing which addresses these shortcomings in turn.
The key outcomes of this ongoing work so far include the introduction of a \textit{relative metric} which limits the cell size dynamically based on the energy of the event (rather than having a fixed maximum cell size applied to events with low and high $p_T$), a study of the individual channels entering the multi-jet merging which dominate the negative event fraction and, an investigation of the advantage of applying cell resampling at the level of weighted events (prior to unweighting).

%% file: tools/reweighting/reweighting_main.tex
\section{Reducing Negative Weights in Simulations and Simulation Resampling}
\label{sec:reweight}
\emph{\noindent J.~R.~Andersen, M.~V.~Garzelli, A.~Maier, V.~Mikuni, S.~Plätzer}
\vspace*{0.5em}

\begin{figure}[t]
    \centering
        \includegraphics[width=.45\textwidth]{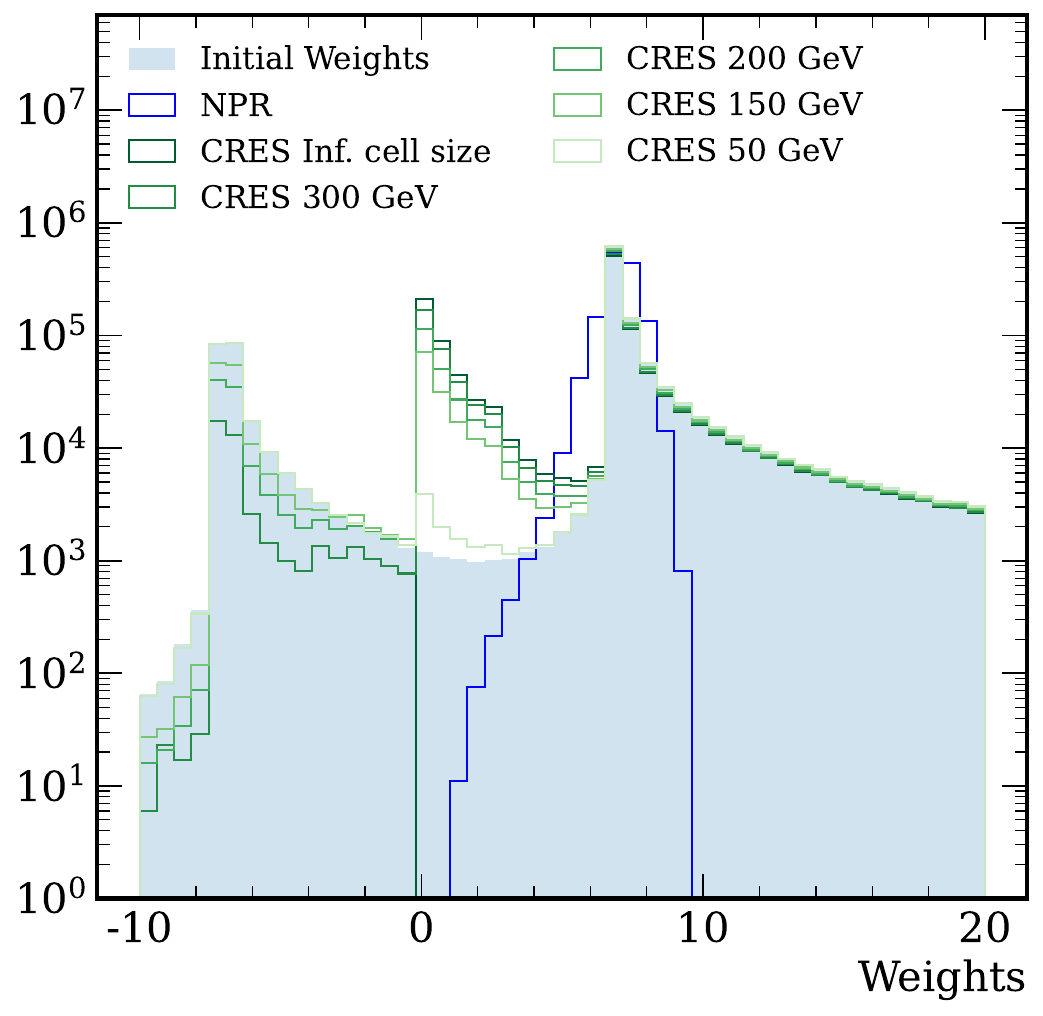}
    \caption{Initial sample weights from the \ttbb sample (blue filled), after NPR (blue line), and after CRES using different cell sizes (different shades of green).}
    \label{fig:npr_weights}
\end{figure}

\noindent Weighted samples are commonly used in collider physics to ensure ensembles of observations have the correct statistical properties. These weights can be determined already at initial stages of the simulation chain, where the average weights in a given phase space point is physically relevant. In particular, at next-to-leading order (NLO) corrections, quantum loops can yield negative weights to individual samples, often requiring large simulation datasets to be stored to achieve statistical power. Instead one may want to reweight samples by reducing the fraction of negative weights such that all kinematic distributions are unchanged to the statistical accuracy of the original sample and the sample size can be reduced, thus reducing the computational cost of the following steps of the simulation chain. During dedicated sessions at Les Houches, different strategies were discussed. The first method was introduced in~\cite{Andersen:2021mvw,Andersen:2023cku} using a cell resampling (CRES) method to modify the weights such that, within a neighborhood, the fraction of negative weights is reduced. The introduction of a maximum cell size ensures kinematic distributions are unchanged on scales larger than the maximum cell size. A large neighborhood leads to lower fractions of negative weights but may also lead to less accurate results. Alternatively, machine learning for neural positive reweighting (NPR)~\cite{Nachman:2020fff} can be used to learn an approximation of the reweighting function that results in positive weights in addition to preserving the statistical uncertainty of the original set with negative weights. To exemplify the methodology, a sample consisting of \ttbb events with an additional emission were simulated using \textsc{PowHel}~\cite{Bevilacqua:2017cru} interfaced with \powhegbox~\cite{Alioli:2010xd} and \textsc{Helac-1} loop~\cite{Bevilacqua:2011xh} with NNPDF3.0 PDF set~\cite{NNPDF:2017mvq}. The distribution of the weights obtained by NPR and CRES with different cell sizes is shown in Figure~\ref{fig:npr_weights}.

\begin{figure}[t]
    \centering
        \includegraphics[width=.24\textwidth]{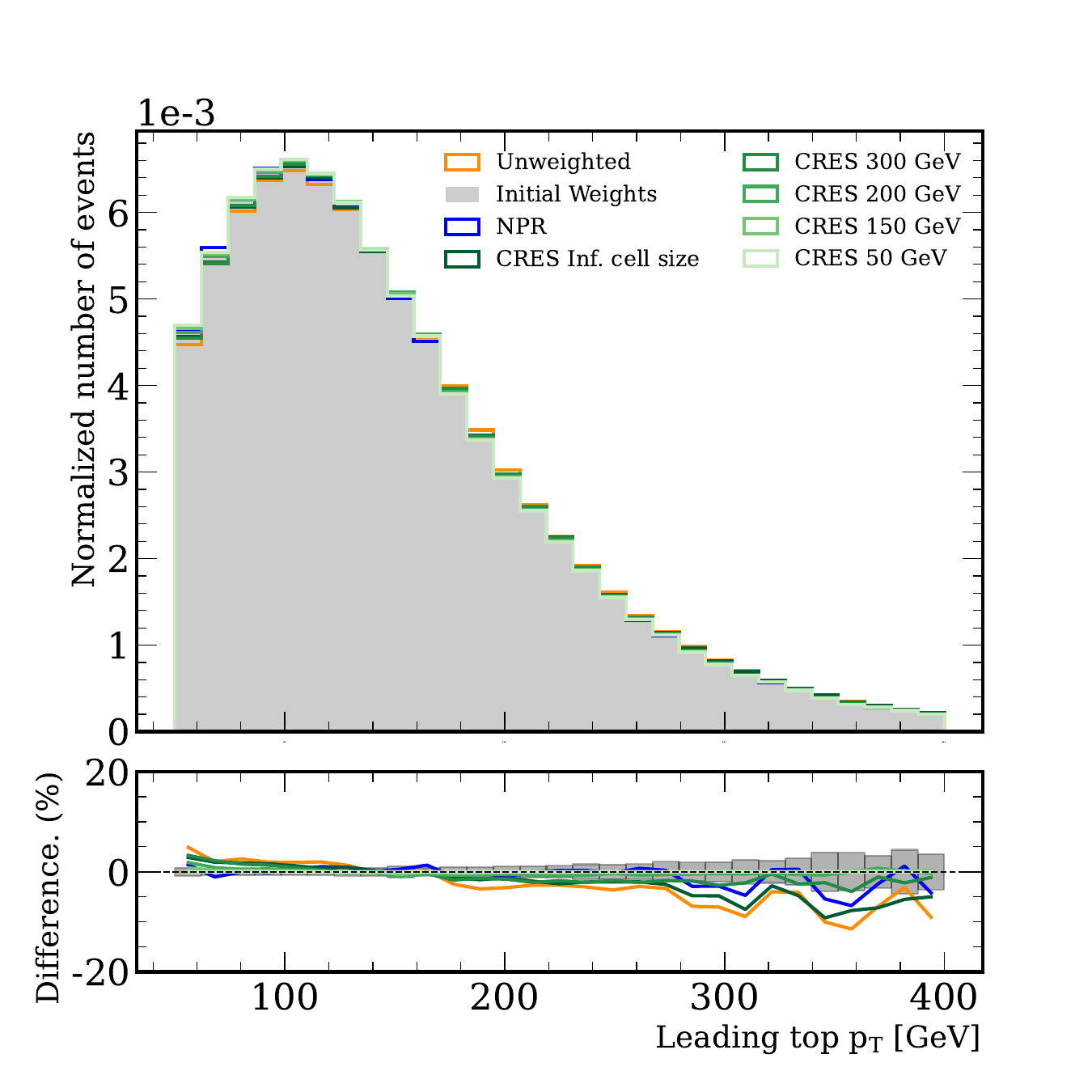}
        \includegraphics[width=.24\textwidth]{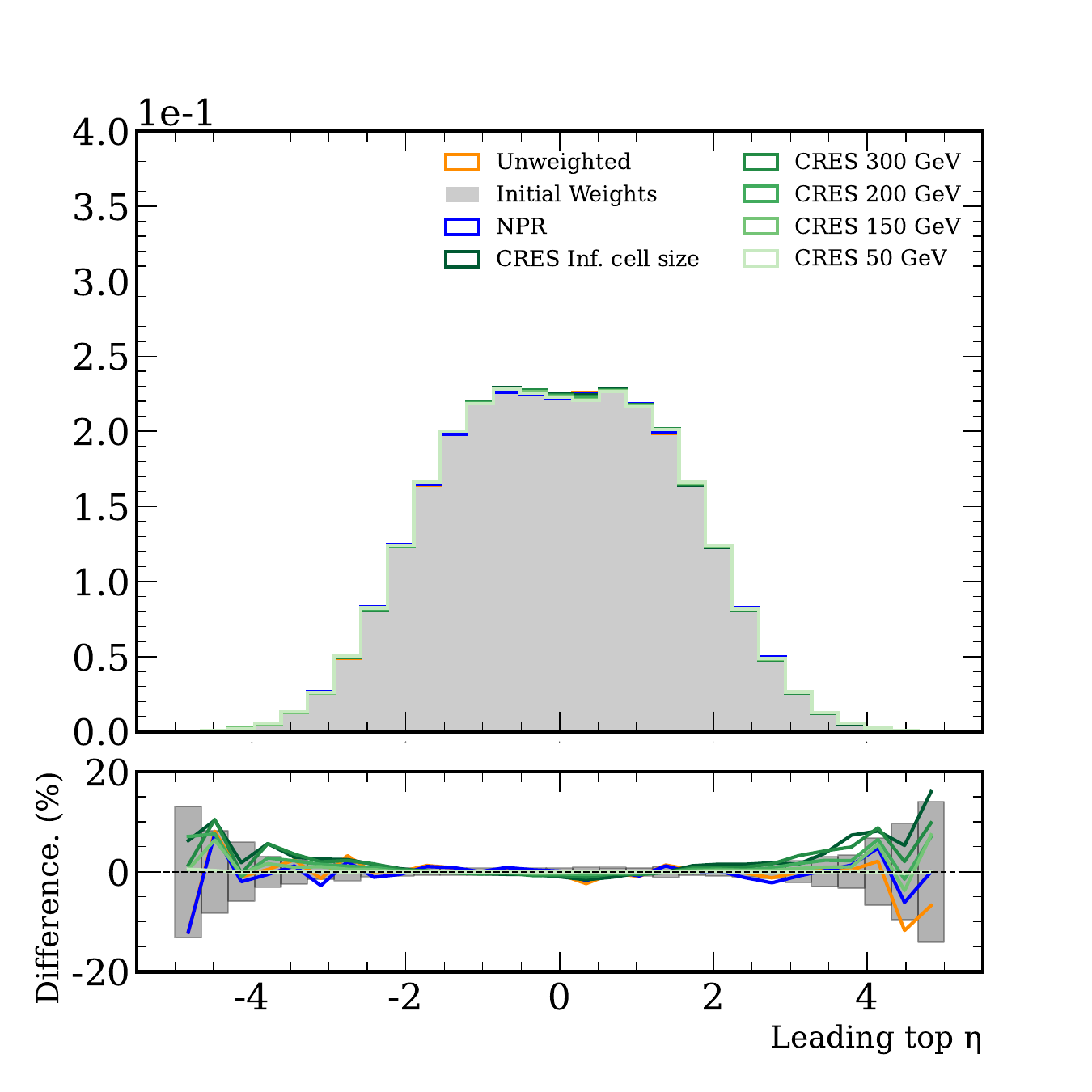}
        \includegraphics[width=.24\textwidth]{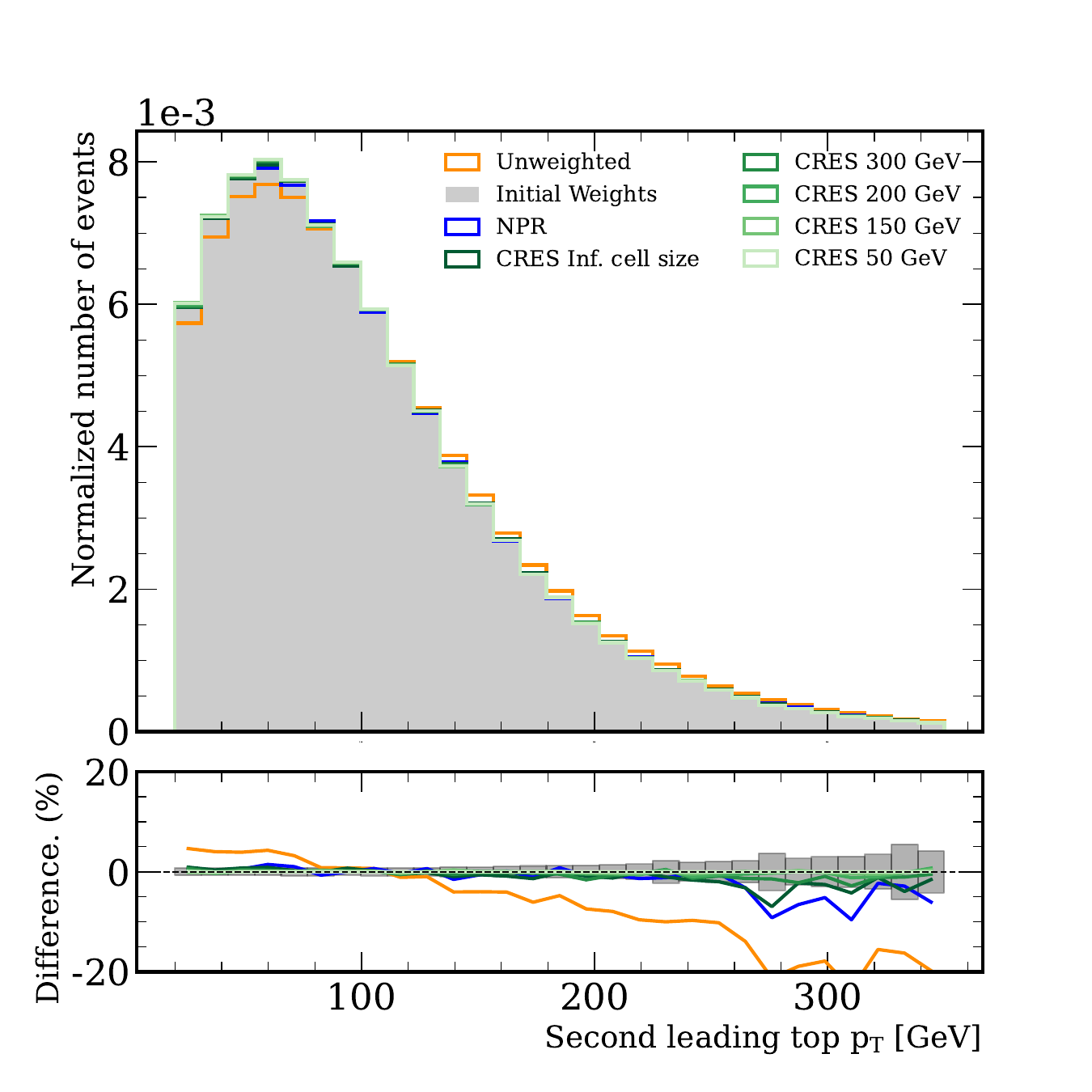}
        \includegraphics[width=.24\textwidth]{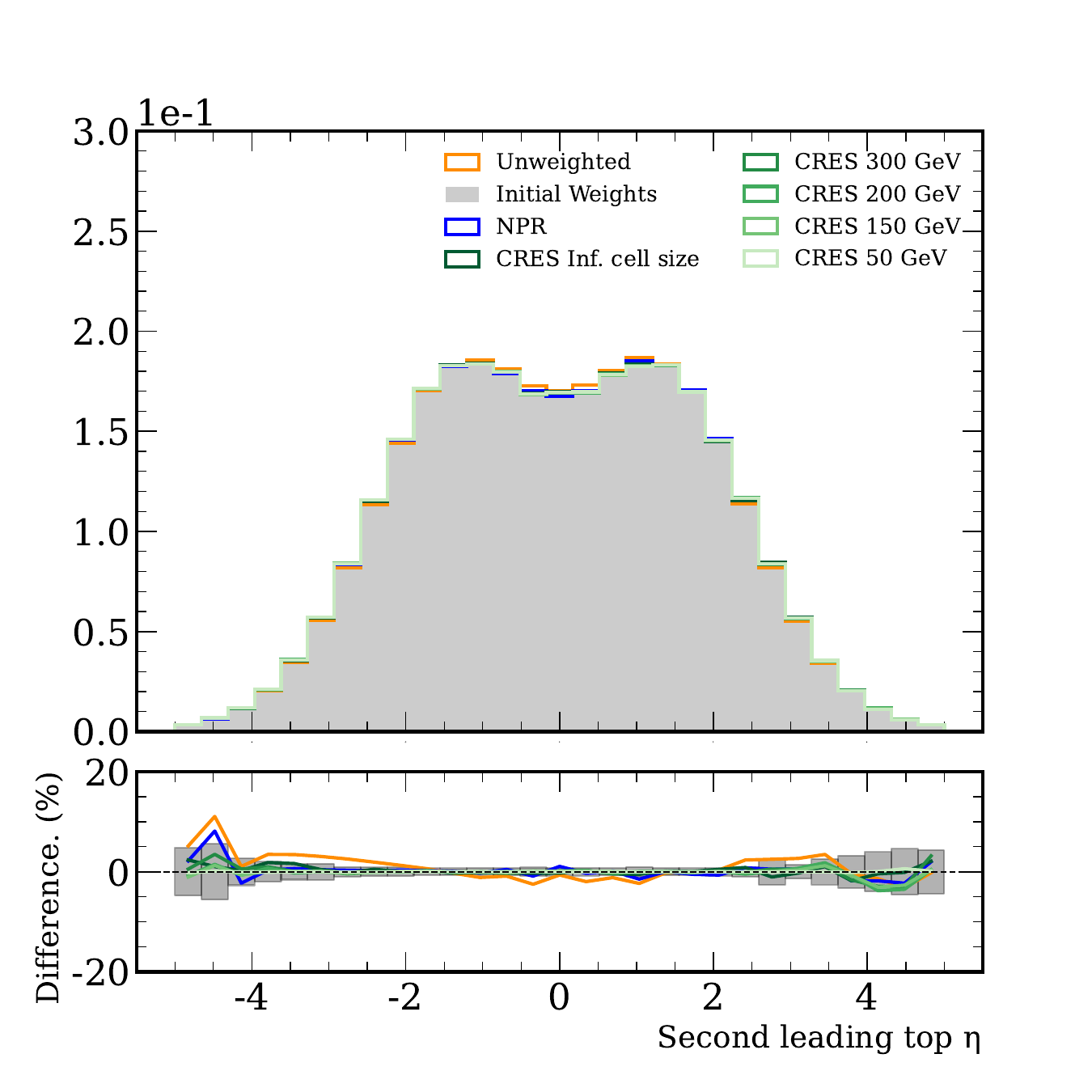}
        \includegraphics[width=.24\textwidth]{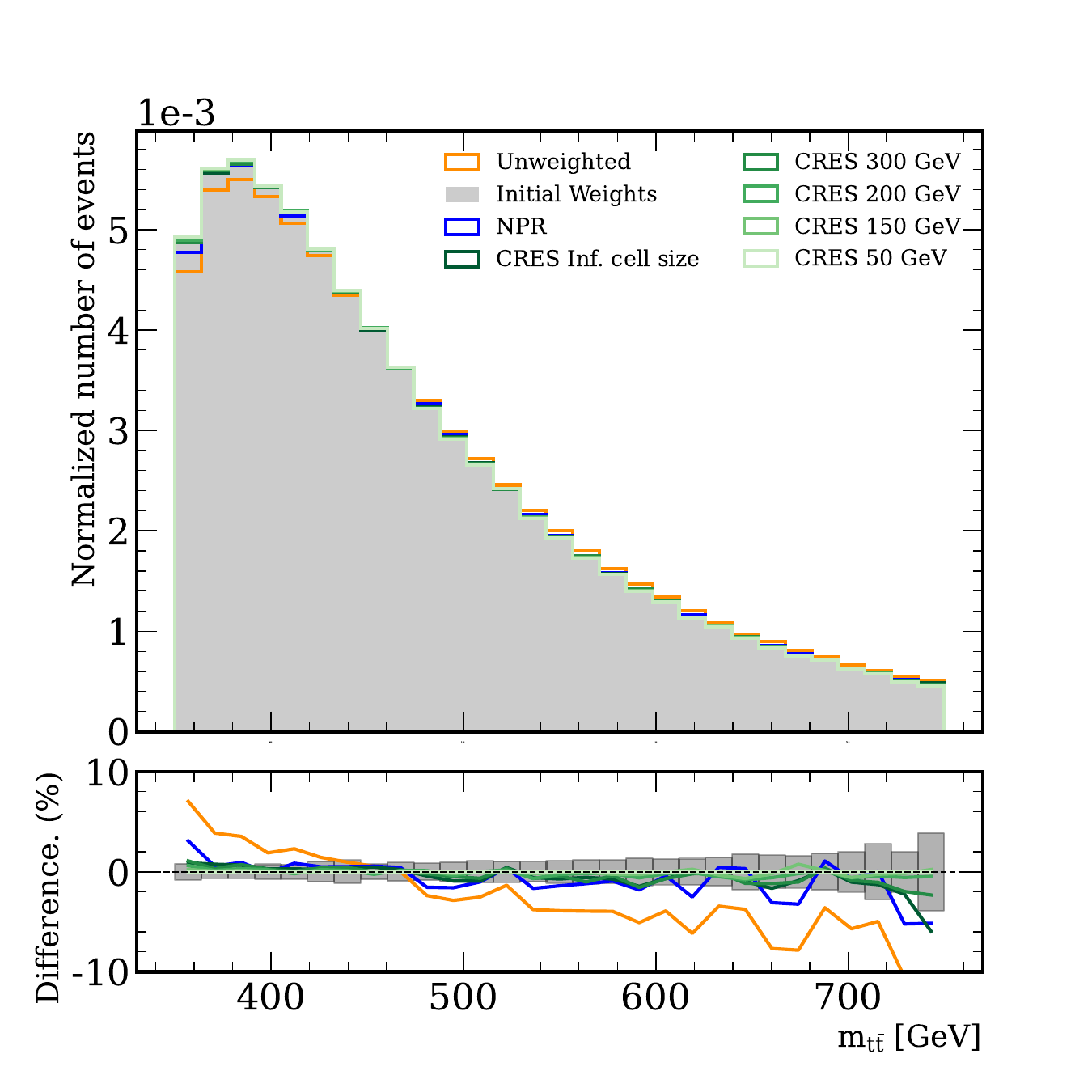}
        \includegraphics[width=.24\textwidth]{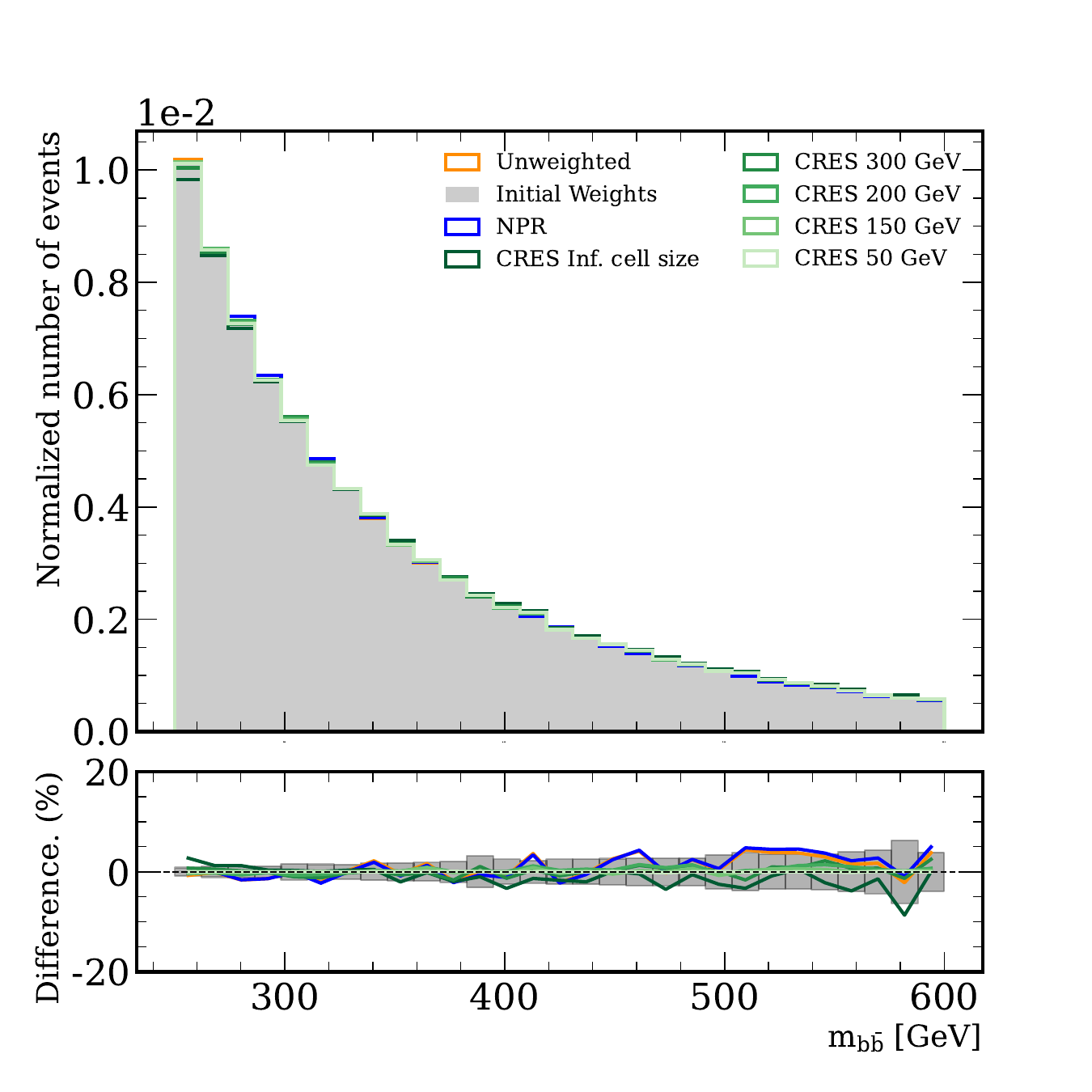}
        \includegraphics[width=.24\textwidth]{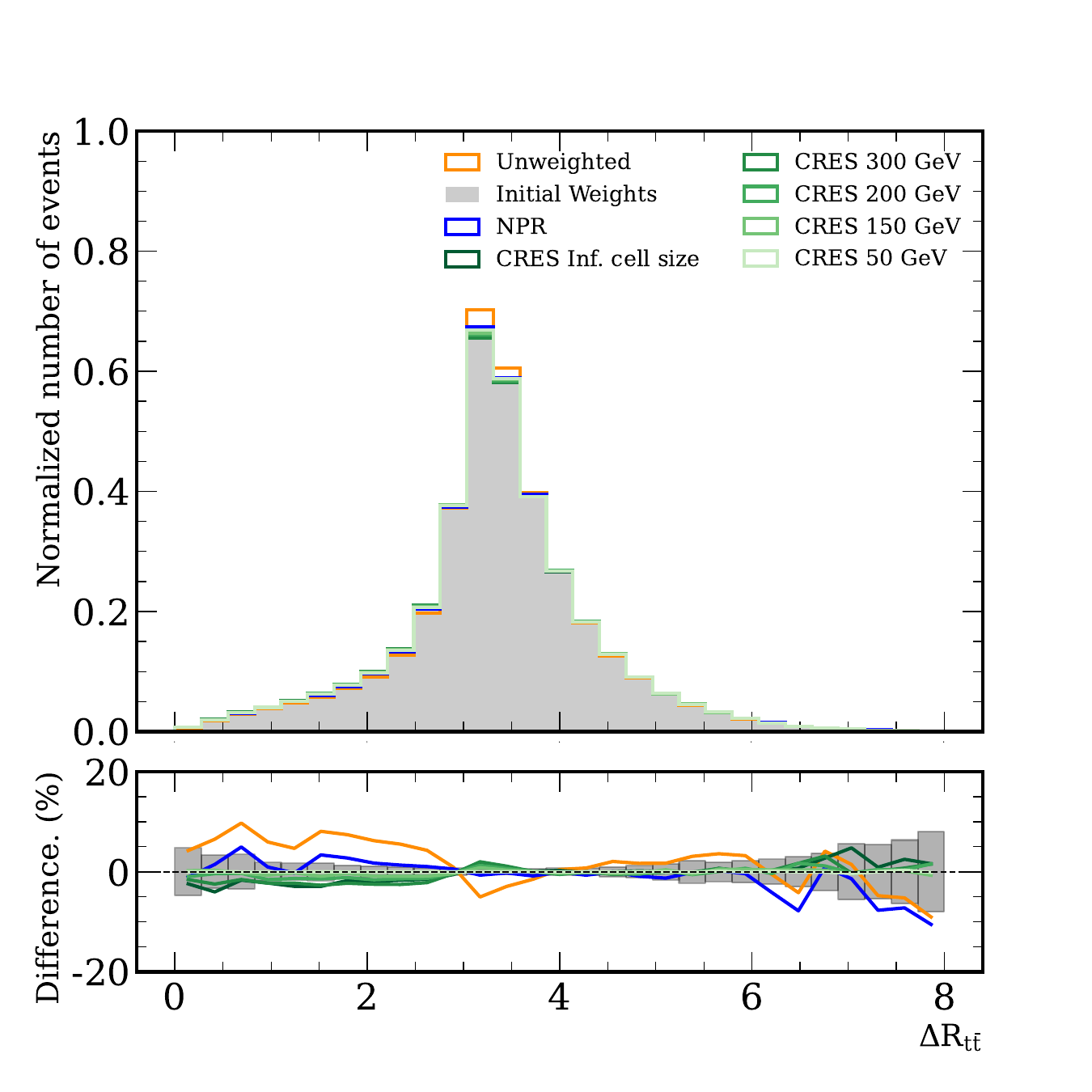}
        \includegraphics[width=.24\textwidth]{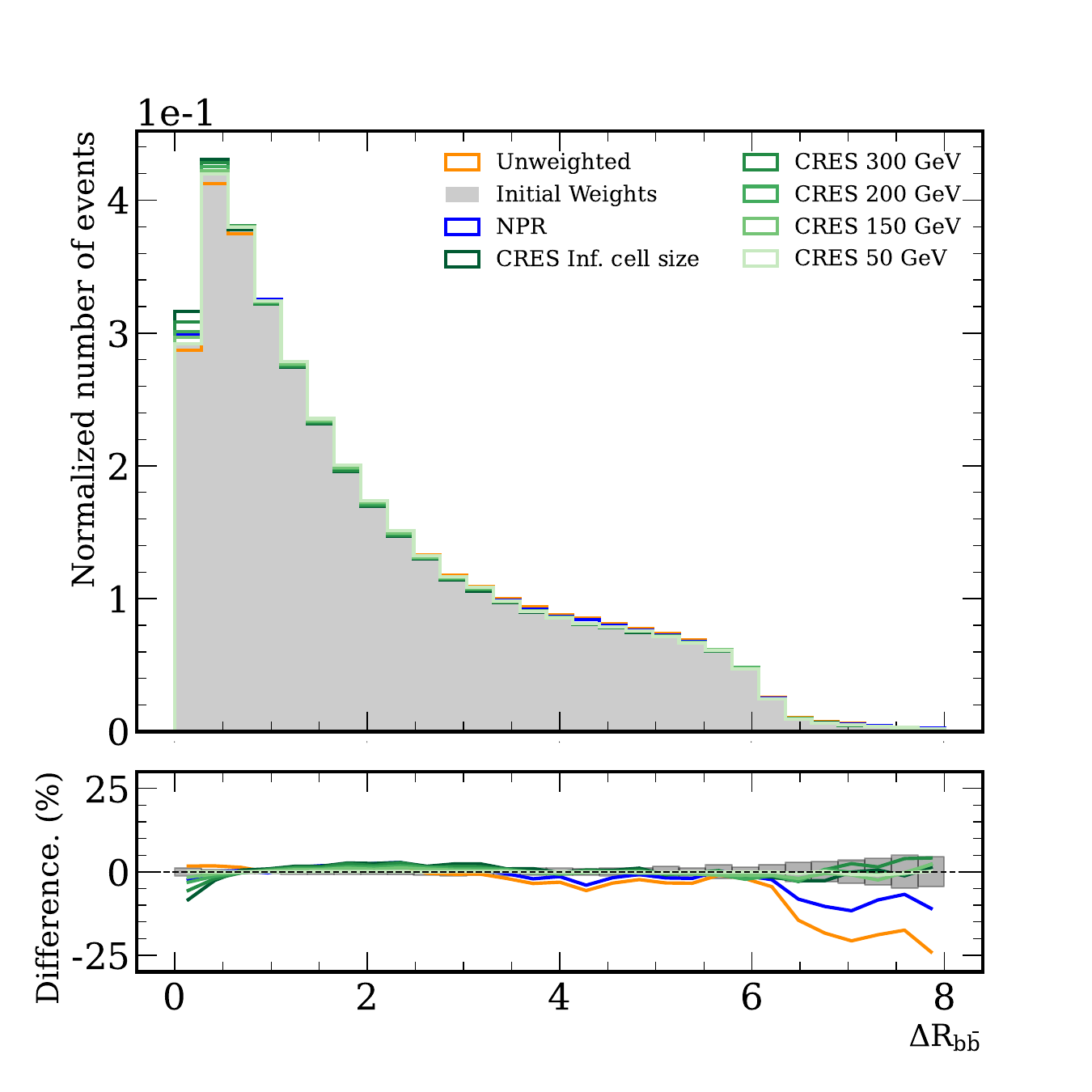}
    \caption{Distributions of multiple kinematic distributions for both top and b quarks. Results without any weights applied (orange), after NPR (blue) and after CRES with different parameters (different shades of green) are compared with the initial distributions carrying negative weights. Percentual difference between each method and original samples are shown in the ratio. Shaded regions in the difference represent the statistical uncertainty of the initial data.}
    \label{fig:npr_kinematic}
\end{figure}

The initial \ttbb sample has 17\% of the weights negative. After reweighting, NPR and CRES with infinite cell size both lead to a sample with only positive weights. Varying the cell size in CRES leads to different fractions of negative weights remaining in the dataset, ranging from 3\% with cell size of 300 GeV to 16.8\% for a cell size of 50 GeV. Both methods are able to leverage the high dimensional feature space, using as inputs the kinematic information from both top quarks, additional b-quarks, and additional emission when resolved. The result is a set of weights valid for any function of these observables. We illustrate this property by verifying multiple kinematic distributions of the system, shown in Figure~\ref{fig:npr_kinematic}. All methods show a better agreement with the initial distribution compared to the unweighted data. NPR results are similar to the CRES results with an infinite maximum cell size, showing that both methods are compatible in the limit when all negative weights are removed. Reducing the cell size improves the accuracy of CRES at the cost of remaining negative weights in the sample. Notice that for applications at LHC, many millions of samples are produced for each physics process, leading to more accurate reweighting for both NPR and CRES.

Finally, during discussions it was pointed out that, in some cases, defining a reweighting function may lead to unstable results. This is the case for samples with high variance of weights. In particular, for parton showering simulations,  the resulting multiplicative weights from the parton shower evolution can lead to a large weight spread. An alternative approach is to use a resampling algorithm~\cite{Olsson:2019wvr}. During resampling, observations are selected in proportion to the event weights, while an interleaved resampling algorithm can be used to reduce the variance of the weight distribution and improving the statistical convergence for parton shower algorithms. 

%% file: tools/unfolding_ML/unfolding_ML_main.tex
\section{Unbinned Unfolding using Machine Learning}
\label{sec:omnifold}
\emph{\noindent M.~Donegà, K. Köneke, V.~Mikuni}
\vspace*{0.5em}

\noindent Unfolding in collider physics refers to the task of correcting physics observables for detector effects. Unfolded observables can be directly compared with different theory predictions, thus accelerating the feedback between experimental results and theory developments. Classical unfolding methods use histograms to unfold physics observables in a small number of dimensions. These methods are often based on inverting a response matrix that describes the migrations between observables due to detector effects. The requirement of binned data in the form of histograms can sometimes be restrictive, as combinations between multiple experiments or including multiple observables in the unfolding procedure require additional attention. Machine learning has the potential to enable the unfolding of multiple observables in a high-dimensional space. In particular, the \textsc{OmniFold}~\cite{Andreassen:2019cjw,Andreassen:2021zzk} method has been applied to multiple studies using hadronic final states using data from the H1~\cite{H1:2021wkz,H1prelim-22-031,H1:2023fzk,H1prelim-21-031}, LHCb~\cite{LHCb:2022rky}, CMS~\cite{Komiske:2022vxg}, and STAR~\cite{Song:2023sxb} Collaborations. During discussions at Les Houches, technical details of \textsc{OmniFold} were discussed, resulting in recommendations and future prospects that unbinned unfolding can achieve a broader impact.  Our recommendations are:
\begin{itemize}
    \item  Always verify the detector acceptance to ensure unfolded observables are within the detector coverage: Even though the measurement is unbinned, results are often reported in the form of histograms. From the unfolded results, a covariance matrix can be determined based on the binning choice and such binning should also be consistent with the detector resolution used to measure the data. 
    \item Unfold a larger region of the phase-space compared to the fiducial region of interest: Acceptance and efficiency effects lead to additional extrapolation uncertainties for unfolded results. Acceptance effects, or the determination of unfolded observables that are outside of the detector acceptance, are important uncertainty sources that are challenging to estimate. On the other hand, efficiency effects due to reconstructed events outside of the fiducial region of interest only lead to sub-optimal use of information, since measured data outside the fiducial region is ignored. For this reason, we recommend to train \textsc{OmniFold} using a larger fiducial, where the complete data information can be included. Since the fiducial selection is not directly used during unfolding, a more restrictive fiducial region can be selected after, reducing the extrapolation uncertainty but still benefiting from the full data.
\end{itemize}
During discussion with the Higgs Physics group, we identified different research directions that would benefit from unbinned unfolding. In particular, multi-differential cross section measurements in the simplified template cross sections (STXS) framework~\cite{LHCHiggsCrossSectionWorkingGroup:2016ypw,Andersen:2016qtm} face the challenge of striking a balance between defining measurement intervals that are granular enough to reduce theoretical uncertainties but inclusive enough to be measured with precision. With unbinned unfolding, the results for each kinematic bin are independent from the unfolding procedure, allowing one to study, with the measured data, the sensitivity obtained for a given choice of kinematic selection.

%% file: tools/MC4SM/MC4SM_main.tex
\section{Machine Learning for Standard Model Measurements}
\emph{\noindent V.~Mikuni, M.~Pellen}
\vspace*{0.5em}

\noindent Machine learning has the capability to extend and improve current methodologies used to extract Standard Model parameters from collider physics measurements. Getting insight into the electroweak-symmetry-breaking mechanism is one of the main endeavors of the LHC Experiment.
To that end, extracting longitudinal modes of weak bosons in scattering processes is a key probe.
In a recent work~\cite{Grossi:2023fqq}, a general method was proposed to tag longitudinally-polarised events on an event-by-event basis. The method relies on ratios of theoretical amplitudes and uses wide neural networks to approximate the ratios, using only as inputs experimentally accessible information. During a dedicated session with the Standard Model phenomenology group, recent results demonstrating the feasibility of this new approach were discussed. Results were presented considering the production of a Z boson in association with a jet at the LHC, both at leading order and in the presence of parton-shower effects.

%% file: tools/WDR/WDR_main.tex
\section{MC Weight Derivatives and Weight Derivative Regression}
\label{sec:wdr}
\emph{\noindent M.~Donegà, V.~Mikuni, A.~Valassi}
\vspace*{0.5em}

\noindent In parameter measurements at colliders, MC reweighting is a well-established technique at least since LEP times, where it was used for measuring the W boson mass~\cite{aleph:1998mw} and anomalous couplings~\cite{lep2:1996tgc,delphi:1998tgc}. With some limitations, this is also applicable to hadron colliders~\cite{gainer:2014rew}, even with QCD NLO accuracy~\cite{mattelaer:2016rew}. In MC reweighting, a single sample is generated at a reference value of a theory parameter and is processed through detector simulation and event reconstruction. A posteriori, event weights are changed based on the event-by-event matrix element ratio between a different value of the parameter and the reference one. Although careful validation is needed, this method offers two key advantages over generating separate MC samples at different parameter values: first, it is computationally cheaper, since only one MC sample goes through detector simulation and event reconstruction; second, it provides a more powerful description of the dependence of experimental distributions on the parameter to be measured, because the variation with the parameter is computed event-by-event rather than bin-by-bin in a given distribution, and is thus less affected by MC statistical errors.

An important quantity encapsulating the event-by-event parameter dependence of MC weights is their derivative with respect to the parameter to be measured. Computing event-by-event weight derivatives at generation time and storing them in LHE files may be particularly interesting for two reasons, as described in a recent paper~\cite{valassi:2020wdr} which was discussed again in Les Houches. First of all, weight derivatives can be used to compute the minimum statistical error that could be achieved from a data sample of given luminosity in an ideal measurement of the parameter, using a detector with ideal acceptance and resolution. In other words, weight derivatives can be used to compute the full Fisher information on the measured parameter that collecting a data sample of given luminosity theoretically offers. Weight derivatives also make it possible to interpret the loss of information and the higher statistical errors achieved in a real measurement, in terms of the experimentally limited acceptance and resolution and of background contamination.

The second interest of weight derivatives is that they can be used for Machine Learning optimizations of the analysis method. The ideal measurement of a parameter, mentioned above, is simply one where the MC-truth weight derivative of each event is precisely known, and where the parameter is measured by the one-dimensional fit of its distribution. This quantity, however, is not an experimental observable, and can only be approximately computed from the existing measured properties of each event. A common way to do this is the Optimal Observables~\cite{atwood:1992oo,davier:1993oo,diehl:1994oo} (OO) method, where the functional dependence of the MC-truth derivative on some MC-truth event properties is known a priori, and an ``optimal observable'' is computed by applying this functional dependency to detector-level event properties. An alternative approach is building a ML regressor for the MC-truth weight derivative in each event, based on the observable properties of each event: this essentially represents the unfolding of this single MC-truth variable. While technically challenging, this ``Weight Derivative Regression'' (WDR) approach is affected by detector effects in a different way than Optimal Observables, and could in principle achieve a better statistical resolution on the measured parameter.

To date, WDR has only be studied on simple toy models, but it has not been attempted on any concrete HEP measurement using real or simulated data. Because of the dependency of weight derivatives on the reference value of the parameter where the derivatives are computed, WDR (like OO before it) seems more suitable to measuring couplings rather than masses, and might be interesting in particular for EFT studies at LHC or Higgs coupling studies at FCC. While the original paper only focuses on measurements of weight derivatives with respect to a single parameter, this framework could partly be extended to simultaneous measurements of multiple parameters at the same time, like many EFT fits. The Les Houches workshop provided a very useful environment to discuss the potential application of weight derivatives and WDR in existing and planned analyses. Some follow-up discussions have already happened after Les Houches and might eventually lead to concrete studies in the future.

%% file: tools/interfaces/interfaces_main.tex
\section{Revisiting event generator, matrix element interfaces and file formats}
\label{sec:interfaces}
\emph{\noindent S.~Plätzer summarizing community discussions and efforts}
\vspace*{0.5em}

\noindent The inter-operability of different matrix element providers and Monte Carlo event generation frameworks is of utmost importance for the field and has culminated in many Les Houches accords, which specify the corresponding interfaces. This includes the Les Houches Event File format (LHEF)~\cite{Alwall:2006yp} to pass events from a fixed-order perturbative calculation to parton shower simulations, or the Binoth Les Houches Accord (BLHA)~\cite{Binoth:2010xt,Alioli:2013nda}, enabling a runtime interface to one-loop matrix element providers, making their results directly accessible in matching or merging frameworks, such as \herwig~\cite{Bellm:2015jjp,Bellm:2019zci}, \pythia~\cite{Sjostrand:2014zea,Bierlich:2022pfr}, \sherpa~\cite{Gleisberg:2008ta,Sherpa:2019gpd}, \powhegbox~\cite{Alioli:2010xd} and \mgamcnlo~\cite{Alwall:2014hca}. 

HepMC \cite{Dobbs:2001ck,Buckley:2019xhk}, on the other hand, has become the de-facto standard or storing and communicating events at the end of event generator simulations. The highly parallelized and/or GPU accelerated calculations of hard matrix elements described in Sec.~\ref{sec:gpu} do, however, require more efficient I/O solutions. A first proposal based on the HDF5 format~\cite{Hoche:2019flt,Bothmann:2023ozs} proved to be an efficient option, and promises to keep up with the rapidly evolving requirements in the exascale computing era. Similar methods would be desirable to interface parton showers and hadronization, with the role of multi-parton interactions still to be clarified. This discussion has been accelerated by the emergence of new and computationally improved parton shower algorithms, such as \cite{DeAngelis:2020rvq,Seymour:2024fmq}. New formats or old proposals \cite{Butterworth:2010ym} for runtime interfaces, or their effective in-house implementations within event generator frameworks, will continue to be discussed and condensed into a new set of Les Houches accords. The underlying physics assumptions, which can differ between the different shower and hadronization models, may play an important role for any potential new standard. Similar considerations apply to interfaces between amplitudes, parton showers, and hadronization not only in light of amplitude evolution algorithms \cite{DeAngelis:2020rvq} but also for a consistent communication of spin~\cite{Richardson:2018pvo} or colour correlations~\cite{Platzer:2012np,Platzer:2018pmd}.

%% file: tools/reproducibility/reproducibility.tex
\section{Containerisation and reproducibility}
\vspace{0.5em}

\noindent One common theme during discussions was the need to make collaboration easier, especially between theorists and experimentalists. One effort in this direction is to push the need for containerisation and reproducibility. 

In practise this can take the shape of generators, run cards and rivet in Docker images. Also existing configurations can be stored through YODA~\cite{Buckley:2023xqh} files, with the possibility to have these uploaded to HepData~\cite{Maguire:2017ypu}. This is already done in some cases for measurements from the experiments but could also be useful when generator teams make their own validations of e.g. a released production version. 
Another benefit of recommended Docker images is that new users (experimentalists) can quickly set up and run generators themselves. One can imagine for example a Docker-based workflow on e.g. lxplus could be created, maintained and documented to lower barrier to entry for MC studies. A first attempt at such a framework was put together.

The reproduciblity of setups in the experiments can also help make their use cases available for benchmarking, including full chain in experiments, for resource profiling.